# Metasurface Energy Harvesters: State-of-the-Art Designs and Their Potential for Energy Sustainable Reconfigurable Intelligent Surfaces

Alireza Ghaneizadeh, *Member, IEEE*, Panagiotis Gavriilidis, *Graduate Student Member, IEEE*, Mojtaba Joodaki, *Senior Member, IEEE*, and George C. Alexandropoulos, *Senior Member, IEEE*

*Abstract*—Metasurface Energy Harvesters (MEHs) have emerged as a prominent enabler of highly efficient Radio Frequency (RF) energy harvesters. This survey delves into the fundamentals of the MEH technology, providing a comprehensive overview of their working principle, unit cell designs and prototypes over various frequency bands, as well as state-of-the-art modes of operation. Inspired by the recent academic and industrial interest on Reconfigurable Intelligent Surfaces (RISs) for the upcoming sixth-Generation (6G) of wireless networks, we study the interplay between this technology and MEHs aiming for energy sustainable RISs power by metasurface-based RF energy harvesting. We present a novel hybrid unit cell design capable of simultaneous energy harvesting and $1$-bit tunable reflection whose dual-functional response is validated via full-wave simulations. Then, we conduct a comparative collection of real-world measurements for ambient RF power levels and power consumption budgets of reflective RISs to unveil the potential for a self-sustainable RIS via ambient RF energy harvesting. The paper is concluded with an elaborative discussion on open design challenges and future research directions for MEHs and energy sustainable hybrid RISs.

## I. INTRODUCTION

Future wireless communication networks are expected to serve massive numbers of low-powered sensors and devices following the Internet of Things (IoT) as well as the Integrated Sensing And Communications (ISAC) paradigms [1]–[4]. This expectation will pave the way for the concept of smart cities which are envisioned to be widely deployed offering sophisticated applications, such as automated transportation, urban security, user localization, and environmental tracking [5]–[7]. However, two key challenges will arise for their realization: *i*) energy supply to power millions of devices and keep them connected to the internet, while *ii*) offering users and smart devices with ubiquitous connectivity, ultra-low data transfer delay, and high Quality of Service (QoS) under harsh propagation environments [8]. Conventional ways to tackle the former include periodic battery replacements and wiring, but both do not seem a viable solution regarding massive scale deployment. As for the latter, typical ways to serve more users and satisfy strigent QoS constraints involve increasing the transmit power of Base Stations (BSs) (which deteriorates the energy efficiency of the network), moving to new frequency bands (which is a limited system resource), and deploying more BSs (i.e., densifying the network) [9].

Smart Radio Environments (SREs) emerge as a revolutionary wireless networking paradigm capable of over-the-air signal programmability in a cost- and energy-efficient manner, and for various wireless objectives [10]–[13]. SREs are networks where the optimization of their analog and digital components for communications, localization, sensing, and their integration merge seamlessly with the aid of the technology of Reconfigurable Intelligent Surfaces (RISs) [14], [15]. In the SRE paradigm, multiple and possibly densely deployed RISs serve to extend the propagation medium to a reconfigurable and optimizable medium, thanks to their ability to alter their reflection properties [16], [17]. In this manner, connectivity issues can be overcome irrespective of the wireless environment conditions.

An RIS is an artificial planar surface, which consists of low-powered, or even passive, metamaterials that under the appropriate actuation mechanism can manipulate incoming Electro-Magnetic (EM) waves [18]. In fact, the RIS elements have been shown to yield changes in all possible characteristics of the EM field, such as its phase, amplitude, polarization, and frequency [19], [20]. A major advantage of this technology is that it eliminates the need for power-hungry Radio Frequency (RF) chains, since interactions with the impinging signals take place entirely in the analog domain. This distinct feature of RISs renders them especially attractive for large-scale deployments [21]–[23].

Although, as already pointed out, RISs are low-power consuming and can be optimized to enhance the overall energy

The work of A. Ghaneizadeh and M. Joodaki has been funded by the Deutsche Forschungsgemeinschaft (DFG, German Research Foundation)-grant number 511400365, JO 1413/3-1. The work of P. Gavriilidis and G. C. Alexandropoulos has been supported by the SNS JU project TERRAMETA under the European Union's Horizon Europe research and innovation programme under Grant Agreement No 101097101, including top-up funding by UKRI under the UK government's Horizon Europe funding guarantee.

A. Ghaneizadeh and M. Joodaki are with the School of Computer Science and Engineering, Constructor University, 28759, Bremen, Germany (e-mails: aghaneizadeh@constructor.university and mjoodaki@constructor.university).

P. Gavriilidis is with the Department of Informatics and Telecommunications, National and Kapodistrian University of Athens, 15784 Athens, Greece. (e-mail: pangavr@di.uoa.gr).

G. C. Alexandropoulos is with the Department of Informatics and Telecommunications, National and Kapodistrian University of Athens, 15784 Athens, Greece and with the Department of Electrical and Computer Engineering, University of Illinois Chicago, Chicago, IL 60601, USA. (e-mail: alexandg@di.uoa.gr).

Corresponding authors: George C. Alexandropoulos and Mojtaba Joodaki.

efficiency of the network, a pressing question remains: how can RISs, along with all the SRE devices, be powered? The large cumulative power required to power all devices in an SRE has rendered traditional methods (battery storage and wiring) cost-ineffective, considering their size, lifetime, cost, and monitoring management. To this end, energy harvesting from surrounding RF signals and Wireless Power Transfer (WPT) have arisen as key approaches to fulfil this challenging requirement [24]–[44]. The Metasurface Energy Harvester (MEH) paradigm has emerged as a promising technological enabler to supply the SRE energy demands from ambient RF and microwave radiation [25]. It was showcased in [25] that the attractive physical properties of their constitutive metamaterials can enable highly efficient energy harvesters, by optimizing the current distribution over the entire metasurface to create a matched impedance network with respect to the environmental impedance. Very recently, an analytical investigation of an energy autonomous RIS via RF energy harvesting was presented in [41], [42].

In this paper, motivated by the increasing interest in RISs, and their diverse hardware architectures and operation modes [45], as well as in their resulting SREs for the upcoming sixth-Generation (6G) of wireless networks [46], we review MEHs and elaborate on the MEH potential for implementing next generation sustainable RISs. The primary contributions and innovations in this review paper in comparison with recent MEH studies (e.g., [25], [30], [47]–[49]) and proposals on energy autonomous RISs (i.e., [41]–[43]) are summarized as follows:

- A comprehensive elucidation of the working principle of MEHs is presented utilizing both the transformer and circuit models.
- The advantageous features of 2D isotropic MEHs are highlighted for enhancing angular stability in ambient RF energy harvesting.
- We introduce a novel RIS hardware architecture comprising 1-bit hybrid meta-atoms, each being capable to perform both RF energy harvesting and two-state tunable reflections.
- We present a comprehensive collection of ambient RF power levels from real-world measurements to identify the most prominent frequency bands for energy harvesting with metasurfaces and provide a rough estimate of the capabilities of ambient RF energy harvesting.
- An exhaustive study of the power consumption of state-of-the-art RIS prototypes, based on experimental measurements, is presented.
- Two novel RF energy harvesting policies, namely the frequency and polarization splitting policies, are presented.
- A thorough scalability analysis that capitalizes on experimental power consumption measurements for the publicly available RIS prototypes and real-world ambient RF power measurements is presented.
- We provide insights regarding various systematic particularities that need to be considered regarding energy autonomous RISs via ambient RF harvesting, such as the frequency bands of operation, the size of RISs, the hardware components comprising RISs, and their energy harvesting principle.
- We list future directions regarding energy consumption minimization and unit cell designs for emerging metamasurface-based architectures and their interplay with energy harvesting.

We believe that the above contributions, not only enhance the conceptual understanding of state-of-the-art MEH structures, but also showcase their promising features for next generation energy sustainable RISs.

## II. THE MEH WORKING PRINCIPLE

We focus on metasurface energy harvesting in the far-field region, avoiding limitations associated with near-field WPT. Despite the fact that in the near-field region highly efficient WPT systems can be achieved via beam focusing [50] using large aperture antennas, a fundamental challenge arises due to the requirement for proximity between the transmitter and receiver [51]. In addition, it is crucial to consider various factors and prerequisites within the Fresnel zone, often referred to as the radiating near-field region. One critical aspect of the WPT system in the Fresnel zone is the use of the lens to precisely focus EM energy at a specific focal point where the receiving node is placed [52], while slight misalignment in the focus can lead to significant performance degradation.

Recent research on using metasurfaces in ambient RF energy harvesting applications can be classified into two main approaches with respect to the far-field region [47]. The first approach involves adding the metasurface structure into the rectenna (rectifying antenna) to enhance the performance of the existing receiving antenna. The second approach introduces a concept where the metasurface directly acts as the MEH instead of a traditional receiving antenna [47]. For RF energy harvesting systems, the MEH is one of the most important elements that can significantly improve the overall harvesting performance.

MEHs are inspired by the metamaterial-based perfect absorbers proposed in 2008 [34], [47], [53]. Such absorbers usually have three layers [54]: a subwavelength-thick dielectric substrate layer, a metasurface resonator layer (such as the metal patch, square loop, cross-type resonators or the Split-Ring Resonator (SRR)), and a highly reflective layer (such as the metallic ground layer) [55], [56]. The physical mechanism of metasurface absorbers can be explained through various approaches [55], [57], including the reflection interference model [58]–[60], the Fabry-Perot (FP) cavity model [55], [61], [62], as well as the impedance matching, circuit, and transmission line models [56], [63], [64]. Unlike metasurface perfect absorbers which dissipate the absorbed EM wave energy within the structure or convert it into heat [34], [47], [65]–[67], the main goal of MEHs is to capture the propagated waves in ambient environments and direct them to guided waves, e.g., in a coplanar or microstrip waveguide or a coaxial cable [65], [66], [68]–[87]. In this way, an ideal MEH is one that converts all power of the impinging plane wave into Alternating Current (AC) power, without any re-radiation of the incident wave. Such a MEH will operate as an EM gate for transferring the energy to the MEH's load(s). In addition, a distinct advantage of MEH is that it can act as a low-profile high-gain antenna for receiving and transmitting information-bearing RF signals (i.e., for realizing wireless communications) rather than just energy harvesting, as will be demonstrated in Section III.

### A. Preliminaries on RF Harvesting Efficiency

The efficiency of an MEH is conceptually similar to the antenna's aperture efficiency, which is different from the radiation efficiency. We typically refer to this as the captured efficiency $\eta_{RF-AC}$, denoting the amount of AC power, $P_{AC}$, available at the output terminal port(s) of the MEH, divided by the incident EM power on the MEH panel, $P_{RF}$, i.e., [28], [49], [73], [88]–[90]:

$$\eta_{RF-AC} = \frac{P_{AC}}{P_{RF}} \quad (1)$$

In a full-wave simulation model, $P_{AC}$ is the power delivered to the MEH load, while $P_{RF}$ is the set power of the plane wave incident on the MEH surface [28]. In order to design and optimize an MEH, a full-wave simulation model involves utilizing the unit cell within a two-Dimensional (2-D) infinite array [76]. The simulation process uses periodic boundary conditions and the Floquet ports in commercial full-wave simulators, such as HFSS and CST [76]. In real measurements, $P_{AC}$ is the available power at the terminal of the MEH and $P_{RF}$ can be determined by the original laboratory measurement setting [28]. To properly illuminate the fabricated MEH with a plane wave, the fabricated MEH should be located at the far-field region of the transmitter antenna [28], [76]. Then, the incident power $P_{RF}$, which is available at the cross-sectional area of the MEH, is given by:

$$P_{RF} = W_{in} A \quad (2)$$

where $W_{in}$ is the power density of the incident wave and $A$ is the physical area of the MEH. The former parameter can be estimated as [28], [49], [73], [88]:

$$W_{in} = \frac{P_T G_T}{4\pi R^2} \quad (3)$$

where $G_T$, $P_T$, and $R$ denote the gain and power of the transmitting antenna, and the distance between the MEH and the transmitting antenna, respectively.

It is worth mentioning that in some investigations, instead of the physical MEH area ($A$), an effective area ($A_e$) is used in eq. (2), due to limitations of the manufactured MEH [91]–[93]. A representative example of this case is an MEH where the SubMiniature version A (SMA) connector is only connected to the central unit cell of the MEH, while all other cells are terminated by matched loads [91]. In such a case, the measurement is only made on the central cell due to practical limitations. However, the non-uniform mutual coupling of other parasitic cells has an impact on the measured power [91]. Thus, an effective MEH area needs to be defined to facilitate the comparison of measurement results based on the central cell's effective area with simulation results based on the physical area of the MEH unit cell [76]. Basically, the effective aperture of the receiving antenna, $A_e$, is related to its physical aperture $A$ via the aperture efficiency $\eta_A$, as $\eta_A \triangleq \frac{A_e}{A}$ [75]. Furthermore, the maximum effective aperture of an antenna can be evaluated with respect to its maximum directivity as $A_{e,\max} = D_0 \lambda^2/4\pi$, with $D_0$ being the maximum directivity of the antenna and $\lambda$ the wavelength [75], [94]. For instance, the authors in [82] and [95] modeled the central cell of their fabricated MEH by using an infinitesimal electric dipole, which had the maximum effective area of $3\lambda^2/8\pi$. In such scenarios, the physical area of the unit cell is smaller than the effective apertures of the central cells [94], [96], [97]. Another example is when the finite array of an MEH has a limited number of cells [96] or in a case where an MEH is a curved panel [91].

It is crucial to note that mutual coupling among cells at the MEH's edge differs from that of central cells [93], [98]. Hence, we introduce a factor $\eta_{\text{nuc}}$, which when multiplied by the $\eta_{AC}$ gives $\eta_{\text{RF-AC}} \triangleq \eta_{nuc}\eta_{AC}$, as shown in Fig. 1. This approach allows us to consider the factors contributing to non-uniform coupling among unit cells during the evaluation process of MEH performance [76], [82], [91], [95], such as, for example, the limited number of unit cells or the curvature of the metasurface antenna.

It is finally important to point out that, in conventional rectifying antennas, the energy harvesting efficiency refers to the amount of EM incident power transformed directly to Direct Current (DC) power, $P_{DC}$, which can either be stored or used directly for energy supply [90]. We distinguish between the two efficiencies to avoid confusion, and define the total harvesting efficiency for direct conversion from radiation to DC power as $\eta_{RF-DC}$. The relation between $\eta_{RF-DC}$ and $\eta_{RF-AC}$ is conventionally given as $\eta_{RF-DC} = \eta_{RF-AC} \eta_M \eta_{AC-DC}$ [89], where $\eta_{AC-DC}$ accounts for the rectification efficiency, while $\eta_{RF-AC}$ is the captured efficiency of the MEH, which explains its applicability as a comparison metric for different MEH structures. The parameter $\eta_M$ models the impedance matching losses between the MEH and rectifying circuits. As a result, the total energy harvesting efficiency from RF to DC, which involves the effects of non-uniform mutual coupling between MEH's cells, is given by $\eta_{RF-DC} \triangleq \eta_{nuc} \eta_{AC} \eta_M \eta_{AC-DC}$. Figure 1 depicts the aforedescribed schematic diagram of an energy harvesting system.

### B. MEH Structure

An RF energy harvesting panel is typically composited of sub-wavelength periodic elements located so closely to each other that mutual coupling affects the input impedance of each element [65], [66], [76]–[84], [88], [91], [93], [95]–[97]. The working principle of an MEH is based on the resonant nature of the elements together with their mutual coupling, which allow for manipulation of the incident plane wave and matching the free-space and load impedances within a limited bandwidth. The elements of an MEH typically consist of: *i*) a very low-loss dielectric substrate grounded by a metallic layer; *ii*) a metallic scattering pattern printed on the top surface; and *iii* metallic via(s) embedded within the dielectric substrate and connected to the metallic pattern on the top surface and load(s) on the ground. It should be mentioned that the most common form of the geometrical metallic scattering patterns might be easily found in the Frequency Selective Surface (FSS) literature e.g., the printed meandered line, SRR, Electric-Inductive-Capacitive (ELC), metallic loop and cut-wire, traces, and canonical omega resonators, as well as their complementary instances [65]. In addition, the authors in [99] introduced non-metallic resonators as a new candidate for microwave energy harvesting. More specifically, they exploited the mutual coupling among adjacent dielectric resonators in a metamaterial array to experimentally demonstrate the mentioned ability, i.e., absorbing EM wave energy and converting the wave energy into AC power.

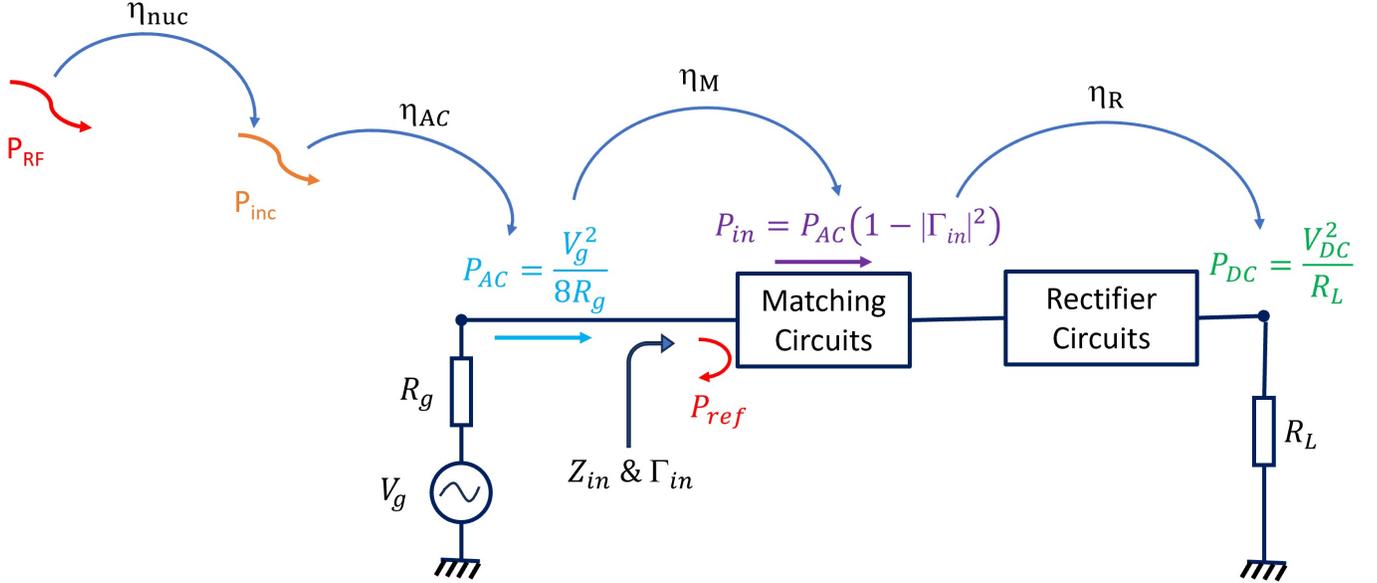

Fig. 1: Schematic illustration of an energy harvesting system. $\eta_{AC}$ is related to the conversion of unguided EM power to the delivered power at the metasurface antenna terminal; $\eta_{nuc}$ includes any effect causing non-uniform coupling among the unit cells, such as a limited number of unit cells and metasurface' panel curvature; $\eta_M$ is the matching circuits' efficiency; $\eta_R$ is the rectifier circuits efficiency; $Z_{in}$ and $\Gamma_{in}$ are respectively the input impedance and reflection coefficient at the input terminals of the matching circuits model; $P_{ref}$ denotes the reflected power due to potential impedance mismatches; $V_g$ is the induced voltage by the incident plane wave impinging on the MEH; and $R_L$ and $R_g$ are the load and MEH impedances, respectively.

*C. Circuit Modeling*

MEHs based on circuit modeling can be easily simulated by EM full-wave simulators and implemented by the inexpensive Printed-Circuit Board (PCB) manufacturing technology. For example, in [82], an MEH with a near-unity capturing efficiency ($\eta_{RF-AC}$) based on an array of ELC resonators has been designed and implemented for a normal incident plane wave at 3 GHz. This example shows that the permittivity and permeability of an MEH are approximately equal to the free-space ones at the resonance frequency. In other words, the reactive part of the incident power can be stored within the MEH resonators, while its real part can be consumed in the conjugately matched loads, similar to what happens in a Resistor-Inductor-Capacitor (RLC) circuit operating at its resonance frequency. In [76], an equivalent circuit model for an MEH cell was proposed. This cell consisted of a patch that functioned as the resonant element on the top layer, while being connected to the ground plane through loaded vias. In Fig. 2, the spatial arrangement of each lumped circuit element is demonstrated on both the top and bottom layers, as well as within the substrate of this MEH cell [76]. Figure 2e showcases the frequency responses of both the circuit and the full-wave models of MEH. In this figure, port 1 is designated for free space impedance (i.e., $377\,\Omega$), while port 2 represents one of the MEH loads. This type of modeling using an RLC circuit expedites the understanding of MEH behavior across a broad spectrum of frequencies, compared to time-consuming full-wave simulations. Furthermore, such a circuit model serves as a matching circuit, making it easier to optimize the MEH architecture to meet specific requirements.

It is finally noted that, for a better conceptual understanding of the RF harvesting mechanism of MEHs, simple ideal circuit models are crucial in mimicking how such metasurfaces work, providing insights on the physical mechanism behind metasurface-based RF energy harvesting, even when the relevant metasurface structures become more complicated.

*D. Transformer Models*

Regarding the conservation of energy principle, each cell of a lossless MEH at its resonance frequency behaves as an ideal transformer with a turns ratio different than unity [65]. Basically, the transformer consists of primary and secondary conductive coil windings around the transformer core. The physical interpretation behind the transformer operation capitalizes on Faraday's law (i.e., magnetic induction). Although an MEH has no conductive winding turns, each MEH cell has a loaded-via whose location distance from the cell's center is proportional to the number of the secondary winding turns [65, Fig. 7]. For example, each MEH cell was modeled by a transformer in [65]. The components of this cell are almost identical to most other MEH cells except for the top metallic pattern layer, including the metallic via(s) and a very low loss grounded-substrate. In this model, if the loaded-via is located at the cell's center (i.e., cells become symmetric), the spatial variation of the induced magnetic flux cannot be sensed by the loaded-via, while the MEH is illuminated by an incident plane wave at the resonance frequency [65], [84]. This symmetric property of cells causes an equal amount of magnetic flux (time-harmonic) movements inside the ultra-thin dielectric substrate from both sides of the via of the cell, as shown in Fig. 3a. However, as shown in Fig. 3b, if the loaded-via is located off-center, the magnetic flux starts rotating in a circular path around the via inside the substrate to produce the curl of magnetic field based on the Ampere–Maxwell law. In order to maximize the energy delivery to the desired load(s) of the MEH, the via's location needs to be optimized, which is inspired by the operational equation of the transformer. Consequently, MEH operates as an energy and impedance transformer between a medium and a load [65].

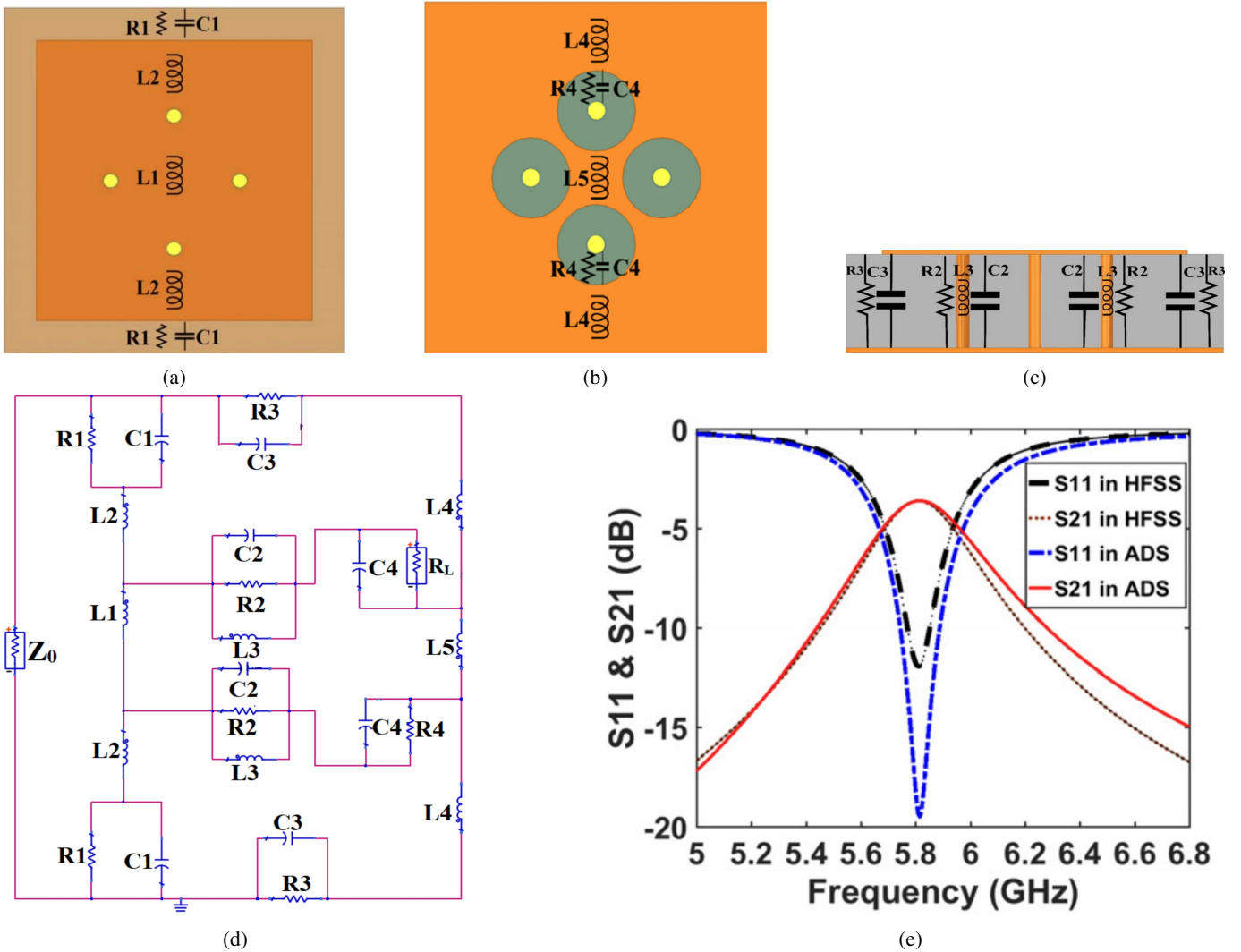

Fig. 2: A schematic view of the MEH unit cell from [76]: (a) the top view; (b) the bottom view; (c) a side view; (d) an equivalent circuit model; and (e) a comparison of simulation results obtained from circuit modeling and from full-wave simulations.

### III. Overview Of MEH Designs

In this section, we provide an overview of the available MEH structures for various bandwidths and frequency bands, emphasizing on their design and fabrication approaches, achieved energy harvesting efficiency, and application scenarios.

#### A. The MEH Pillars

The first MEH array was proposed in 2012 as a new EM energy collector operating in the microwave regime. It was based on the strong resonance and coupling between densely placed MEH cells [83]. In this study, the planar MEH was actually a single-layer energy harvesting array consisting of 81 sub-wavelength SRR cells printed on a dielectric substrate, which was simulated by a full-wave EM simulator [83]. A resistor was also placed in the split of each SRR cell as a load. Later, the authors in [94] aimed to experimentally measure the RF-to-DC efficiency of an MEH based on SRRs. To that end, a Greinacher rectifier circuit was embedded within each SRR cell, with a load resistance of $70\,\Omega$. In order to observe the energy harvesting capabilities of the system, the authors implemented a $5 \times 1$ array of the designed SRRs at the frequency of 900 MHz and placed them in an open Transverse Electro-Magnetic (TEM) waveguide [94]. In 2013, the concept of using the building blocks of metamaterials to harvest microwave energy was extended into the Infra-Red (IR) regime [100]. The authors in [100] proposed a new mechanism for channelling the IR waves from the SRR array to the desired load through a microstrip transmission line. It is noted that most solar energy harvesters are based on the photovoltaic effect [101], [102], which is a quantum phenomenon dealing with the particle nature of light [103], while, interestingly, the IR MEH proposed in [100] exploited the wave nature of IR radiation for RF energy harvesting. For this metasurface, the equivalent circuit model of each MEH cell acts as a matching network between the load and free space impedances due to the surface's ability to tune the effective permeability and permittivity [65], [82], [83].

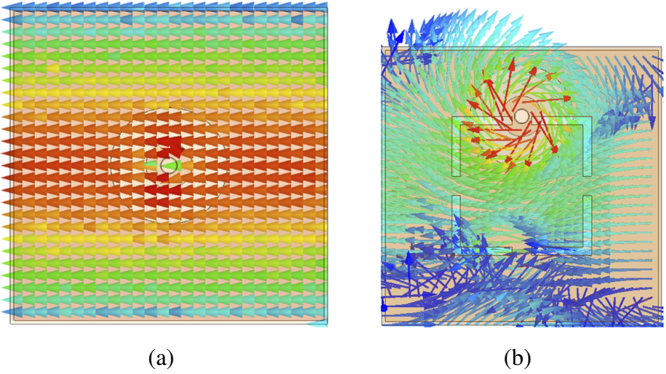

(a) (b)

Fig. 3: Distribution of the magnetic field inside the dielectric substrate of an MEH's cell which is illuminated by a normal incident plane wave; reprinted from [65] and licensed under a Creative Commons By Attribution (CC BY) license. The loaded via is located in: (a) center of the cell; and (b) the off-center of the cell.

An MEH consisting of Ground-Backed Complementary SRR (GCSRR) cells, each with an additional embedded metallic via was proposed in [96]. The via interconnect passing through the substrate wss used for delivering wave energy to the load located between the ground and the via. In this work, a comparison of both bandwidth and efficiency took place between the GCSRR array and a microstrip patch array using the same operating frequency and physical area [96]. Based on empirical results, it was showcased that the GCSRR array has a wider frequency bandwidth and higher efficiency than the microstrip patch array, as shown in Fig. 4 [96]. Consequently, most of the MEH architectures developed later on were mainly advanced in accordance with the aforedescribed typical works, targeting to improve MEH features, such as multi-polarized, multi-band, wide-angle, and wideband receptions [76]. It should be noted that the MEH designers have typically chosen the crowded frequency bands for operation (e.g., Wi-Fi, GSM, LTE). In fact, these frequency bands are desirable for RF energy harvesting because of their significant amount of energy within smart buildings and (semi-)urban regions. Another advantage of these frequencies is that the ambient signal energy is free of charge [65], [76]. The key features and hardware designs of the state-of-the-art MEH structures is presented in the following section.

### B. State-of-the-Art in MEH Designs

As previously mentioned, sub-wavelength ELC cells have been successfully deployed as MEH's resonators, improving the captured efficiency [82]. The simulation and experimental results in that paper showcased that approximately $97\%$ and $93\%$ of the captured efficiency ($\eta_{RF-AC}$) can be respectively achieved. This indicated that almost all incident wave energy was absorbed in the energy harvesting panel and then transferred to the load. This result demonstrated that MEH can act as an EM collector in the rectenna, realizing an effective source to supply energy for low-consumption sensory, and other, devices in a massive IoT smart environment.

*1) Polarization Insensitive and Multiband MEH:* To maximize the harvesting of ambient energy, the design of an MEH with both features of polarization and incident-angle insensitivities is highly desirable, since the ambient RF signals have arbitrary polarization and oblique-angles of incidence [65]. In [88], the authors presented a binary optimization algorithm to optimize the cell shape of a fully automated dual-band MEH. Each cell of the MEH array was composed of a coded-pixelated resonator, four via interconnects, and a grounded substrate loaded by two resistors [88]. The main goal of this work was to maximize the efficiency around two crowded frequency bands. To this end, the proposed MEH was able to achieve more than $90\%$ $\eta_{RF-AC}$ at both 6 GHz and 2.45 GHz for normally incident waves with any linear polarization [88]. In [95], the authors studied a high-efficiency and miniaturized MEH at three operating bands, including the GSM900 (900 MHz), LTE (2.6 GHz), and Wi-Fi (5.7 GHz) frequency bands with polarization-insensitive and wide-angle characteristics. However, this MEH faced the challenging problem of high impedance at the output ports (i.e., impedance matching difficulty of MEH with the rectifying circuits).

On the other hand, multi-band absorption can be achieved by a wideband MEH. For example, a wide-angle and polarization-independent MEH array was proposed in [97] to support a wide operation bandwidth from 6.2 GHz to 21.4 GHz. Each cell of the proposed MEH included a grounded substrate, four metallic vias, four resistors ($50\Omega$), and four metal bars connected to a square-ring resonator [97]. These four symmetric collection ports were employed to improve MEH's performance with respect to the polarization insensitivity characteristic. However, this solution caused the absorbed energy to be delivered to four loads with increased complexity, which renders it inappropriate in real-world energy harvesting applications. In [93], the authors focused on polarization-insensitive and wide-angle MEH with only one terminal port

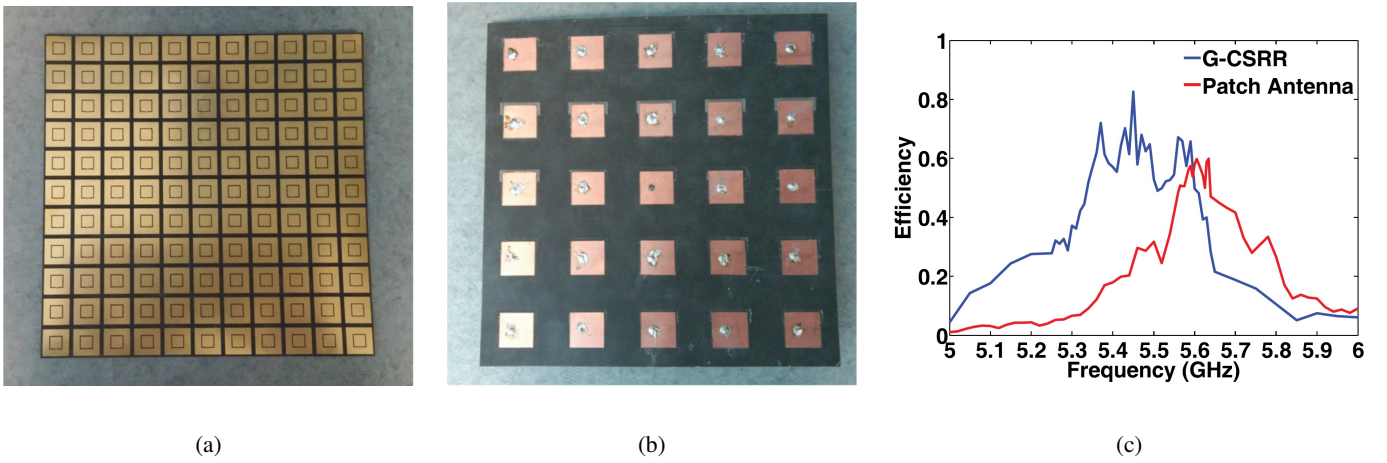

Fig. 4: The top surface of the fabricated: (a) GCSRR MEH panel comprising $11 \times 11$ elements; (b) microstrip patch antenna array with $5 \times 5$ elements; and (c) the energy capturing efficiency of the microstrip patch antenna array and GCSRR MEH measured at their central cell. Reprinted from [96] with the permission of AIP Publishing.

(i.e., $50\Omega$) in each cell. They proposed a simple compact designed MEH array based on a rotating central symmetric resonator structure at $5.8$ GHz.

*2) Conformal MEH:* The previously described MEHs were based on conventional flat and rigid platforms. In [91] published in 2019, the authors designed and fabricated an ultra-thin flexible MEH with an overall thickness of $0.004$ times the incident wavelength at $5.33$ GHz based on the Complementary Quad SRRs (CQSRRs). To develop the conformal capabilities of MEH in real-world harvesting applications, such as wearable smart devices or vehicles, a flexible multi-polarized MEH based on the patch resonators was designed and implemented in [76]. Furthermore, a flexible quasi-metasurface energy harvester, designed to operate at $5.8$ GHz, was proposed, designed, and fabricated in [104] using inkjet-printing technology.

*3) MEH at Extremely High Frequencies:* The role of millimeter-Wave (mmWave), IR, and TeraHertz (THz) technologies is expected to be significant in the upcoming 6G wireless networks, owing to their potential for high data rates and low-latency wireless links [105], [106]. However, despite their potential, there are currently only a few MEHs operating in these spectral regions. For example, a flexible dual-band MEH was proposed at X and V frequency bands, being useful for mmWave applications [84]. In order to improve the sensing capabilities in IR detectors, the authors in [107] proposed to employ SRR cells for harvesting EM energy at IR frequencies. Similar to the integrated nanoantenna load in [108], which was placed among its metallic electrodes ( [108, Fig. 2]), the MEH designed in [107] was integrated with a low band-gap Indium Gallium Arsenide Antimonide (InGaAsSb) Positive-Negative (PN) junction for maximum power transfer to InGaAsSb load placed in the SRR gap. Moreover, a performance analysis of SRR cells was performed in this paper to investigate how different parameters affect its characteristics in the IR frequency range, such as the shape and dimensions of the resonators, type of the metal, place and number of gaps, and loops in SRRs [107]. In 2021, an IR MEH was designed through radar cross-section reduction of an $L$-shaped chiral metasurface array based on a phase cancellation approach [103].

The typical unit cells of MEHs developed for RF energy harvesting are illustrated in Figs. 5a–5i. In Table I, we provide a performance review among the various unit cells employed in state-of-the-art MEH structures.

### C. MEH as a Metasurface Antenna

In [110] and [112], metasurface antennas based on the MEH principle were implemented in an $8 \times 8$ array with periodic unit cells for capturing and radiating the linearly and circularly polarized waves at $3$ GHz and $5.8$ GHz, respectively. An advantage of these planar metasurface structures was their high gain, evaluated at $G = 11.7$ dBi and $14.2$ dBi with full-wave simulations. This gain was achieved in the direction of the normal vector originating from the surface. The dimensions of these structures were $1.2\lambda \times 1.2\lambda \times 0.025\lambda$ and $1.57\lambda \times 1.57\lambda \times 0.04\lambda$ for operation at $3$ GHz and $5.8$ GHz, respectively, with $\lambda$ being the free space resonance wavelength. Their aperture efficiencies ($\eta_A = G\lambda^2/4\pi A$) are $\eta_{A,3\text{GHz}} = 0.817$ and $\eta_{A,5.8\text{GHz}} = 0.849$, respectively; note that both values can be considered nearly ideal since a good aperture efficiency is in the range of $0.5-0.8$ for conventional antenna arrays [113]. Furthermore, these architectures follow the microstrip-based feeding network, and to achieve the highest antenna gain in the normal direction of the antenna surface, their radiating cells were fed in-phase due to the uniform surface current distribution on the cells except the ones at the antennas edges. It should be noted that, while the captured efficiency is the most important feature of MEH as a collector, in its design as an antenna, typical antenna features are important, such as impedance matching, gain, polarization, radiation pattern, etc. [110]–[112], [114].

The authors in [109] employed an MEH as an alternative to classic antennas to activate the rectification circuit for providing DC output. Figures 5j and 5k depict the linearly polarized TMA proposed in [110] and the coding circularly polarized TMA designed in [111], respectively. In [115], a wide-angle broadband rectifying metasurface structure for energy harvesting and WPT in the microwave frequency range was designed and fabricated. In [98], the authors designed and implemented a metasurface energy collector with harmonic suppression, aiming to enhance the RF-to-DC conversion efficiency. Notably, this novel approach exhibited insensitivity to variations in the load and input power. In contrast to conventional antenna arrays, the MEH unit cells were so densely placed that in-layer feeding networks became infeasible [109], [110]. It is noted that these additional required layers may introduce more manufacturing errors and losses [98], [109], [110], [112], [116], [117]. In order to convert the radiated power to DC, the integrated co-planar design of the metasurface collector and diodes might be considered as an effective solution for removing the mentioned additional layer [116], [118]. However, it should be noted that, although this single layer of harvesting surface can solve the mentioned issues, the real problem is that it cannot act as an MEH in its typical forms, as shown in Figs. 5j and 5k. This limitation stems from the tight spatial separation between MEH cells, making the implementation of the co-planar feeding network impractical.

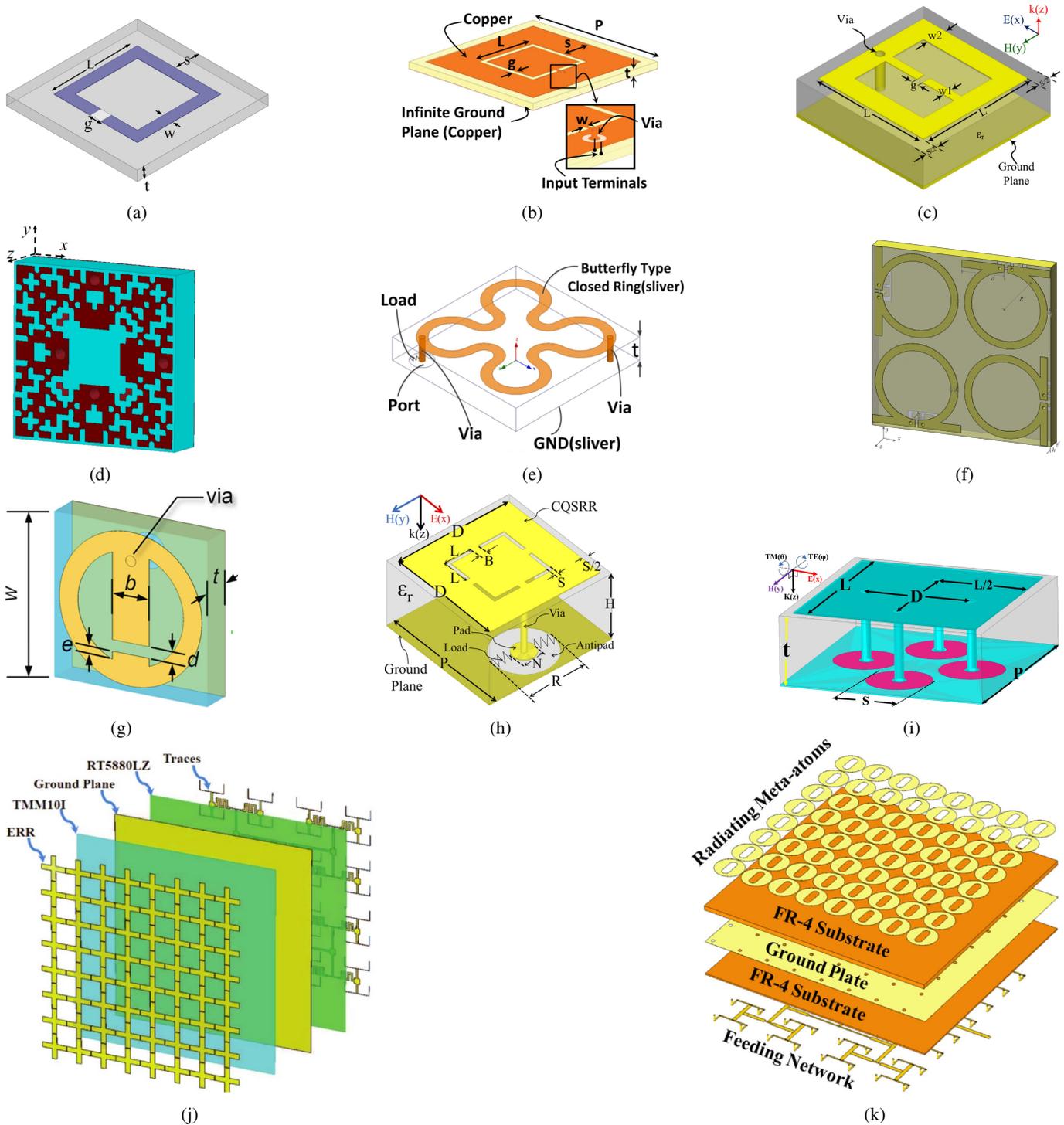

Fig. 5: The schematic view of different MEH cells developed for ambient RF energy harvesting: (a) SRR unit-cell, reprinted from [83] with the permission of AIP Publishing; (b) GCSRR unit-cell, reprinted from [96] with the permission of AIP Publishing; (c) ELC unit-cell, reprinted from [82] with the permission of AIP Publishing; (d) Coded-pixelated unit-cell, reprinted from [88] and licensed under the CC BY license; (e) BCR unit-cell, reprinted from [95] with the permission of AIP Publishing; (f) omega ring resonators, reprinted from [78] and licensed under the CC BY license; (g) metallic mirrored split rings with hollow cylinder, reprinted from [72]; (h) CQSRR unit-cell, reprinted from [91] and licensed under the CC BY license; (i) Patch unit-cell, reprinted with permission from [76]; (j) the layered perspective view of the linearly polarized TMA, reprinted from [110] and licensed under the CC BY license; and (k) the circularly polarized coding metasurface antenna, reproduced with permission from [111] and copyrighted by Wiley-VCH GmbH.

TABLE I: Performance overview of unit cells employed in MEH's top layer.

| Ref. | Freq. (GHz) | No. Bands | No. Via/ Shape of MEH cell | Substrate/ Flexible | Periodicity of cells $/\lambda_0^*$ | Thickness (mm) | Termination impedance of each via | Sensitivity to linear polarization |
|---|---|---|---|---|---|---|---|---|
| [95] | 0.9, 2.6, 5.7 | 3 | 2/BCR | F4B/No | 0.08 | 4 | 2776($\Omega$) | No |
| [32] | 1.75, 3.8, 5.4 | 3 | 4/Four rotating SRR | Rogers RO4003/No | 0.18 | 1.524 | 240($\Omega$) | No |
| [72] | 2.45 | 1 | 1/Mirrored SRR and Cylinder | PTFE/No | 0.16 | 4 | 50($\Omega$) | Yes |
| [73] | 2.45 | 1 | 2/ELC | Rogers RT6006/No | 0.09 | 2.54 | 200($\Omega$) | No |
| [34] | 2.45 | 1 | 1/Face-to-face shared gap SRRs | Rogers 04350/No | 0.12 | 4.572 | 90($\Omega$) | Yes |
| [28] | 2.45, 5.8 | 2 | 2/Modified complementary Jerusalem cross | F4B/No | 0.2 | 4 | 50($\Omega$) | No |
| [88] | 2.45, 6 | 2 | 4/Pixelated cell | Rogers 6006 RTduroid/No | 0.09 | 2.5 | 240($\Omega$) | No |
| [74] | 2.5 | 1 | 1/Square-ring | F4B/No | 0.13 | 3 | 500($\Omega$) | No |
| [109] | 3 | 1 | 1/ERR | Rog. TMM10i/No | 0.15 | 1.542 | 200($\Omega$) | Yes |
| [82] | 3 | 1 | 1/ELC Patch | Rogers TMM10i/No | 0.07 | 2.54 | 82($\Omega$) | Yes |
| [91] | 5.33 | 1 | 1/CQSRR | Rogers RO3010/Yes | 0.13 | 0.254 | 50($\Omega$) | Yes |
| [96] | 5.55 | 1 | 1/G-CSRR | Rogers RT5880/No | 0.34 | 0.79 | 50($\Omega$) | Yes |
| [68] | 5.6 | 1 | 4/WG-CSRR | Rogers RO4003/No | 0.2 | 1.524 | 90.2($\Omega$) | No |
| [65] | 5.77 | 1 | 1/CQSRR | Rogers RO3010/Yes | 0.12 | 0.254 | 50($\Omega$) | No |
| [93] | 5.8 | 1 | 1/roating central resonator | -/No | 0.32 | 0.787 | 50($\Omega$) | No |
| [78] | 5.8 | 1 | 2/Omega Ring | Rogers RO4350/No | 0.16 | 1.524 | 50($\Omega$) | No |
| [98] | 5.8 | 1 | 1/HSMS cell | Rogers 5880/ No | 0.25 | 1 | 50($\Omega$) | Yes |
| [76] | 5.85 | 1 | 4/Patch | Rogers R03010/Yes | 0.14 | 0.254 | 50($\Omega$) | No |
| [97] | 6.2 to 21.4 | *Wide band* | 1/Square-ring with 4 slits | FRB-2/No | 0.19 | 3 | 50($\Omega$) | No |
| [84] | 10.1, 42.86 | 2 | 1/C-Slot Patch | Rogers RO3010/Yes | 0.11 | 0.254 | 20($\Omega$) | Yes |

### D. Rectifying Metasurfaces

The authors in [119] proposed a new energy collector utilizing the cut-wire unit cell integrated with a PN-junction diode within the cell. They also investigated how the diode can activate itself in the presence of a surrounding resonator without requiring any biasing circuit. By adding a resonator to the diode, a suitable voltage difference can be created across the diode. This is because the resonator, triggered by the plane wave radiation, generates a large electric dipole momentum [119]. It is noteworthy that, despite the fact that the low-loss cut-wire cell considerably enhances the electric field around the resonance element, the captured field tends to re-radiate instantly because of the coupling between the electric dipole and free space modes [119]. To address this issue, a narrow gap was introduced at the center of each cell, and a PN-junction diode was inserted within the gap as a nonlinear element. Within this work's simulation-based study, a conductive path throughout all cut-wire cells was shown to be needed to collect DC current from all cells. To establish this path, thin metal wires with high inductance were added between adjacent unit cells. Their high inductance mostly impeded the flow of the surface current among the cut-wire unit cells, and the highest energy conversion efficiency, $\eta_{RF-DC}$, of $50\%$ was achieved at $6.75$ GHz, where the thin-wire integrated diode exhibited resonance [119]. The capability of cut-wire metasurfaces to harvest EM waves has been experimentally studied in [120] at the $2.4$ GHz central frequency. This metasurface structure was constructed using a 3D printing technique, which was essentially a distinct manufacturing process. The measured harvesting efficiency $\eta_{RF-DC}$ reported was around $22\%$ at its peak, which is lower than conventional PCB-printed cut-wire metasurfaces, but their low-cost and easy implementation makes them an attractive choice.

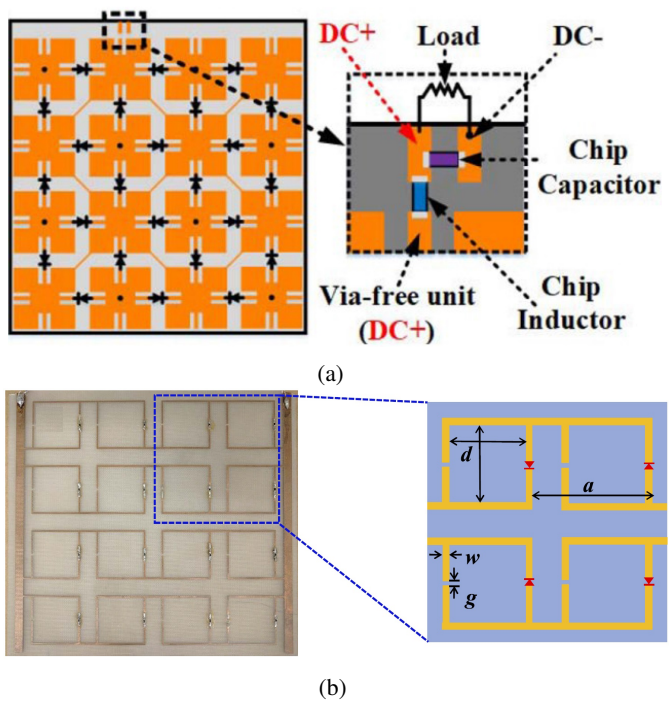

Fig. 6: The layout design model of MEHs whose diodes are embedded in the resonator's gaps: (a) The design of [116]; and (b) The design of [92], reprinted with permission ©Optica Publishing Group.

Recently, in [89] and [121], a new approach was proposed based on periodic FSS to develop scalable FSS-based rectennas that can harvest EM waves in one and two bands, respectively. In this approach, the unit cells are connected together to construct energy transfer channels that accumulate incident power, which is then collected at a terminal and passed to the rectifier circuit [89], [121]. Following this approach, energy loss caused by using multiple diodes instead is mitigated. Furthermore, the designed unit cells enable scalability in the sense that EM wave energy at different input power densities can be harvested. However, the dimensions of the FSS-based resonant cells were relatively large and sensitive to polarization in [89], [121]. To overcome these challenges, a dual-band and wide-angle array based on compact centrosymmetric unit-cells (i.e., polarization-insensitive) for RF energy harnessing in the UMTS-2100 and Wi-Fi bands was proposed in [122]. In [116], a wide-angle dual-band harvesting surface has been designed without the need of a matching network between the MEH and the rectifying diodes, as shown in Fig. 6a. The authors in [116] proposed a new method for co-planar integrating diodes into the MEH top surface, which reduces the complexity of the compact single-layer MEH design. This dual-band harvesting surface was designed at the two crowded frequencies of $5.8$ GHz and $2.4$ GHz, based on a multimode structure which was not sensitive to the linear polarization of the incident wave [116]. Furthermore, a three-band MEH was designed and implemented by arranging a number of SRRs in a nested configuration on the MEH top layer for maximizing the RF-to-DC conversion efficiency [117]. This MEH achieved a capture efficiency of $38.3\%$ at $5.3$ GHz under transverse magnetic polarization and $66.5\%$, $40.6\%$, and $35.6\%$ at $2.4$ GHz, $5.2$ GHz, and $5.8$ GHz under transverse electric polarization, respectively [117].

An MEH based on an FP cavity model of metasurface absorbers [62] to convert the received radio wave into DC power was recently designed and analyzed in [92]. To exploit the FP cavity resonance of the MPA model, the authors employed two-gap SRR cells on a metallic ground plane at a distance of $4$ cm at $900$ MHz ($\approx \lambda/8.5$); in this way, the diodes were placed in the gaps as shown in Fig. 6b [92]. In this architecture, the existence of the ground plane significantly affects the frequency response characteristics of the energy harvesting efficiency. Specifically, with a ground plane present, the efficiency has a peak at $900$ MHz, which is $16$ times bigger than the respective peak without the ground plane, indicating an increase in the quality factor. On the other hand, without the ground plane, the energy efficiency has a wider peak in frequency. Different from the literature, where high-intensity levels were assumed, the intensity levels considered in [92] matched those available from ambient RF sources ($\sim 1\mu W/cm^2$) (for comparison, in [117], the measurement results were obtained when the incident power density was more than $66\mu W/cm^2$). In more detail, the RF-to-DC efficiency achieved with the ground plane present, was measured at $80\%$ for a power density level of $0.4\mu W/cm^2$, and even higher than $100\%$ efficiency was achieved for power density levels larger than $2.6$ $\mu W/cm^2$ [92]. The latter can be attributed to the fact that optimal placement of the ground plane causes an increase in the effective aperture, which exceeds the physical aperture of the antenna (i.e.,$\eta_A > 100\%$) [92].

In another approach [123], a single-layer dual-band scalable array was designed. The $\eta_{RF-DC}$ efficiency of this array at $2.4$ GHz and $12.6$ GHz was $64\%$ and $55\%$, respectively, while the available incident EM power on the array surface was respectively $3$ dBm and $14$ dBm [123]. In [124], the authors proposed a dual-layer miniaturized rectifying metasurface that features a particular arrangement of $8$ asymmetric dipoles in each polarization-insensitive cell. The rectifier diodes on each dipole were conjugate-matched at 2.45 GHz by tuning the length of the dipole arm [124]. In [125], the author proposed and fabricated a new flexible single-layer rectifying metasurface structure. This designed structure exhibited polarization insensitivity and triple-band characteristics, operating at the frequencies 2.9 GHz, 5.2 GHz, and 5.77 GHz [125].

*E. Metasurface Absorbers*

A broad bandwidth polarization-insensitive perfect absorber was proposed in the THz regime in [126], which can be tuned with re-scaling and re-sizing of its geometric configuration. This graphene-based metamaterial absorber can be developed for energy harvesting in the THz spectrum [126]. In [127], a broadband polarization-insensitive metasurface absorber based on wheel resonators was presented. Additionally, in [128], a wideband load-insensitive metasurface absorber was introduced, which utilized symmetrical square-loop ring resonators. Furthermore, [129] presented an ultra-thin dual-layer triple-band flexible metasurface absorber based on six distinct concentric annular rings. These innovative absorber designs could also serve as good candidates for harvesting EM wave energy.

## IV. 2D Isotropic MEH Designs

The exponential increase of IoT devices motivates the development of efficient EM energy collectors to empower them. To this end, MEHs emerge as an enabler for supplying energy to low-powered devices and IoT (e.g., sensors, remote controllers, as well as implantable and wearable electronics), by harvesting either ambient RF power [30] or serving as power collectors from a dedicated source, and consequently, as power providers. This could be done, for example, by mounting MEHs on such devices so that they become self-sustainable energy-wise, thu,s contributing to the sustainability visionary feature of 6G, and beyond, wireless networks.

One question that naturally arises when considering MEHs for energy harvesting is the following: why are MEHs necessary for powering low-power devices if traditional array antennas deployed over a ground plane can also achieve an ideal aperture efficiency? Concerning the weak ambient RF power density, which is typically in the µW range, the answer can be mainly found in the physical footprint of the energy collector [84]. It is first noted that, although most of the low-power IoT devices are designed to be hand-held and/or sufficiently small, typical antenna arrays need inter-element spacing of half-wavelength to avoid destructive mutual coupling. In contrast, the distance between MEH elements is in the scale of deep sub-wavelength, resulting in the miniaturization of metasurfaces and their inherent capability for device-level integration (e.g., in biomedical applications). Hence, due to the restrictive spacing in conventional arrays, fewer antenna elements can be placed in the same physical area compared to MEHs. Therefore, designing a traditional antenna array for powering IoT devices becomes challenging, if not impossible, when considering the restricted size of the device, especially, in the high operating frequency bands.

In practice, however, the diversity of the ambient signals with respect to polarization, bandwidth, frequency, and angles of the incident plane wave leads to limitations in terms of system performance [65]. For example, the direction of arrival of the incident RF signals from the surrounding environment on the MEH structure is random, which causes random fluctuations of the surface wave impedances [76]. The challenge here is the difficulty, in fact, the inability, of the planar MEH to have a variable impedance to be capable to match with the various surface wave impedances of the ambient signals. This represents a fundamental limit on the look-angle of planar MEHs to receive ambient signals isotropically [65], [76]. As aforementioned, there are many articles that deal with enhancing the angular stability of MEHs as well as the traditional array antennas, which is an important feature for the self-adaptability of ambient RF energy harvesting and wireless communications. For example, in 2007, a spherical array antenna was designed in [130], which consisted of 95 sub-array elements, as shown in Fig. 7a. The distinctive advantage of this antenna was its beam scanning capabilities in a complete hemisphere with nearly constant directivity or beam width. Besides, each sub-array included 3 circularly-polarized microstrip patch antennas designed in a conventional way at the center frequency of 9.5 GHz [130]. In 2017, the design concept of a cylindrical metasurface antenna was introduced based on constructive phase interference, which might be highly suitable for wireless communication towers [131]. Figure 7b shows the numerical simulation results of the omnidirectional radiation pattern of this cylindrical metasurface antenna with a maximum gain of 7.55 dBi [131]. The first 2D isotropic flexible MEH was designed and implemented by a small slice of a cylinder in 2019 [91], as depicted in Fig. 7c.

Figure 7d illustrates a flexible MEH design wrapping on a cylindrical structure that can receive ambient plane waves from arbitrary directions in the horizontal plane (i.e., omni-directionally or 2D isotropic) [65]. To meet the needs for this purpose, a cylindrical MEH must be considered whose central axis is perpendicular to the incidence plane [65], [76]. Specifically, for each angular direction in the horizontal plane, there is one MEH cell that can approximately sense waves from that direction as normal incident waves. A worst-case scenario for this prototype is the direction being perpendicular to the edge of a unit cell; this case was studied for an MEH array with $11 \times 11$ elements in [91], with the MEH wrapping around a small slice of a cylinder with a bending radius of 48 mm at 5.33 GHz. Due to the good angular stability of the MEH, the results of [91] showed that the deviation angle of the normal incident wave (the reported $\theta$-angle in Fig. 7c is $101°$ [91]) and other bending effects can be ignored for 5 cells around the central cell. Figure 7d shows this scenario in a simple schematic view [65] with three sections determined to clarify the incidence plane view, represented as follows: 1) an example of flexible MEH cells wrapping around a cylindrical MEH; 2) an example of the ambient incident plane waves to cylindrical MEH with any oblique angle; and 3) an example of arbitrary linear polarization of ambient waves [65].

## V. Energy Harvesting RIS

As previously discussed, in the envisioned 6G wireless networks, it is expected that the number of wireless devices connected to the network will exponentially increase [1]–[4], and the frequency range will expand up to the sub-THz band to facilitate immersive applications as well as to provide the bandwidth resources required for the deployment of the numerous devices. These envisioned networks are faced with key challenges, such as high data rate requirements, peculiarities in high-frequency signal propagation (e.g., increased path loss, molecular absorption, and Line of Sight (LoS) susceptibility), massive connectivity, and energy supply for low-powered IoT devices. To resolve the preceding issues, a concept that has attracted significant research focus involves the use of RISs [14], [15] and the resulting SREs [10], [12], [13]. RISs are reconfigurable metasurfaces capable of altering the properties of impinging EM waves (phase, amplitude, and polarization), which enables the engineering of the propagation environment to enhance various communication metrics. Indicatively, RISs have been shown to achieve increased received signal strength and coverage extension (and consequently, bit error rate or data rate) [12], energy efficiency [18], spatiotemporal focusing [11], secrecy [132], [133], multipath enrichment in LoS conditions [14], localization [17], and ISAC via programmable correlation between communications and sensing subspaces [3]. Specifically, coverage extension is one of their most well known applications, envisioned in modern urban communication networks to overcome blind spots that inevitably arise due to large obstacles [134], [135]. In such scenarios, RISs can be deployed on the building walls, effectively transforming structures that once obstructed signals into low-cost access points that favorably configure communication channels and create virtual LoS paths with increased gain. Recently, new models for these RIS-enabled smart communication channels have been developed, extend-ing beyond stochastic channel modelings, to accurately capture the effect of RISs in the wireless propagation environment [136]–[140].

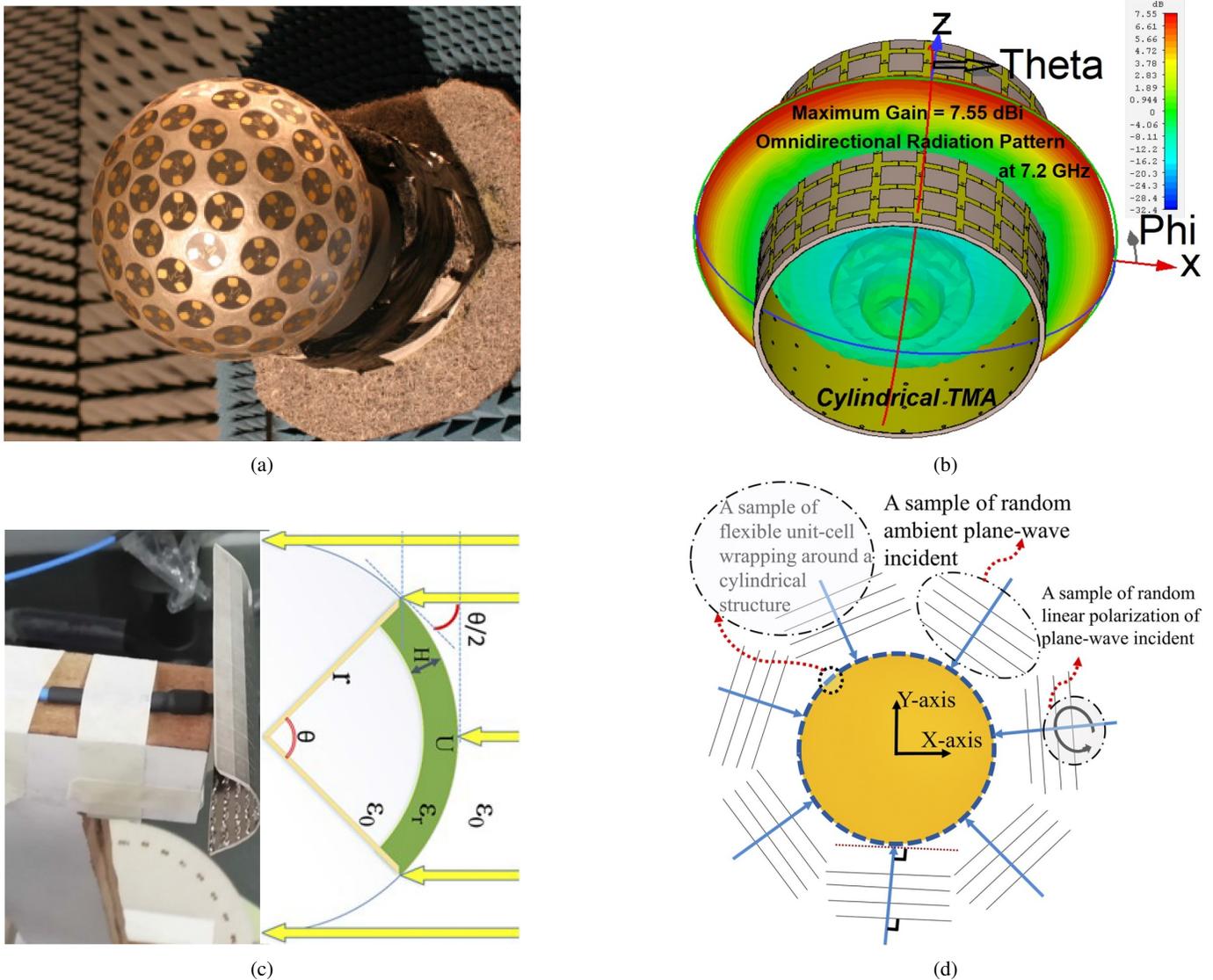

Fig. 7: (a) The fabricated spherical antenna array in [130]; (b) simulation results of the 3D omnidirectional radiation pattern of the conformal metasurface antenna designed in [131]; (c) the fabricated flexible MEH and its schematic side-view model illuminated by a sample of ambient plane-wave in [91], licensed under a CC BY license; and (d) schematic view of flexible MEH wrapping around a cylindrical structure viewing in the incidence-plane in [65], licensed under a CC BY license.

### A. Metamaterial Technologies for RIS

Any artificial planar surface consisting of metamaterial elements (also known as unit cells or meta-atoms) with tunable physical properties, which can be configured via electrical, mechanical, and/or electromechanical stimuli is lately known as an RIS [18]. These actuations are typically controlled by a central processor, such as a Field Programmable Gate Array (FPGA) or a microcontroller [15]. The choice of the metamaterial technology that is implemented in RISs depends on various aspects, such as the frequency range of operation, the beam shaping tunability and accuracy (e.g., fixed anomalous reflection, low-bit tunability, and continuous phase shifting), the required time response, and the available actuation mechanism [106], [141]. Some of the most prominent metamaterial technologies for RISs are listed in the following.

- Micro-Electro-Mechanical Systems (MEMS) and liquid crystals are typically considered for mmWave and THz frequencies. These types of metamaterials are especially attractive due to their ability to retain their working state even when no stimuli is applied (non-volatile switching). However, their development is still in an early stage, and the literature on RIS prototypes is limited [142].
- Microfluids are another promising technology for RISs operating in the mmWave and THz bands, boasting continuous phase shifting and low power consumption. Nonetheless, their fabrication cost is expensive and their mechanical actuation mechanism is typically problemetic when $\mu$s time of response is needed [141].
- RF-switches enabled by Complementary Metal Oxide Semiconductor (CMOS) technology [143] are more mature technologies compared to the aforementioned. They are typically triggered with electrical stimuli and have low unit cell cost and power consumption. In addition, they are capable of EM absorption on top of reconfigurable reflections. Their main disadvantage is that they cannot scale to THz frequencies [142].
- Graphene is considered as an up-and-coming metamaterial technology for operation at the THz-IR spectrum, realizing continuously tunable reflections by dynamic control of its conductivity via electrical stimuli [19]. However, graphene is still an immature technology, and large-scale deployment, as required in the considered large aperture antennas, is still an open challenge.
- Positive-Intrinsic-Negative (PIN) and varactor diodes are the most established technologies for current RIS prototypes [144]–[146] due to their low cost and easy integration in the unit cell. These metamaterials typically work in the MHz up to few GHz frequencies, not being able, however, to reach the THz spectrum. Their main disadvantages are the following. Regarding PIN diodes, the power consumption of each unit cell is an issue. On top of that, their power requirements depend on their working state, i.e., ON/OFF, which can prove problematic due to the instable energy supply needed. Moreover, for multi-bit tunability, multiple PIN diodes have to be inserted in the unit cell, which makes unit cell integration difficult increasing also the range of potential power supply needed. As for varactor diodes, although the unit cell power consumption is negligible, the driving circuitry, and mainly the level regulators, consume a lot of energy, rendering the scaling to large surfaces challenging [144].

Although RISs are less power consuming compared to conventional fully digital antenna arrays, they do not come without a cost, especially when they are supposed to consist of numerous metamaterial elements. The need to power thousands of diodes, varactors, or transistors within an RIS is a significant issue. The conventional solutions involving ordinary batteries require periodic replacements, and traditional wiring incurs high control and installation costs. An innovative alternative for RISs would comprise a hybrid meta-atom being capable to simultaneously harness EM wave energy and offer tunable reflections. This approach will have the potential to enable the control of EM wave energy, thereby, further enhancing the implementation of RIS. Very recently, the authors in [41], [42] investigated analytically the performance of an energy autonomous RIS design intended to realize programmable wireless environments through RF energy harvesting. Also, in [43], the concept of a self-powered RIS via a dedicated power beacon was experimentally demonstrated. It is noted that the unit element design for energy autonomous RISs was not addressed in [41], [42] and the design proposed in [43] contained a PIN diode mounted on the front side of a hybrid meta-atom. Since a rectifier circuit is perpendicularly connected to a metal feed pattern etched on the back side of each block of four meta-atoms, the designed element therein was not a low profile hybrid energy harvesting and reflecting meta-atom. Furthermore, this approach implies that the energy harvesting and reflection functionalities can only be achieved in orthogonal polarizations [43].

### B. Proposed energy harvesting RIS design

In this section, we describe the proposed 1-bit hybrid RIS design for the far-field region which is capable of controlling EM wave propagation and performing RF energy harvesting simultaneously. Our hybrid metasurface integrates an MEH into its structure to serve as the power supply, thus, realizing an energy sustainable RIS.

We adopt the MEH cell designed by Ramahi's group in 2015 [82], where four individual cells combine to form a supercell, as depicted in Fig. 8. This supercell incorporates a central metallic via that connects the top surface of all four cells via four diodes to the ground, which can be controlled through biasing lines and diodes. Each of the proposed supercells has two distinct states: one where the diodes function as short circuits (the ON state in Fig. 8b), and the other where the diodes are open (the OFF state in Fig. 8a), effectively disconnecting the via and the top surface of the four cells. To evaluate the captured efficiency of the proposed design, full-wave simulations were carried out for both states.

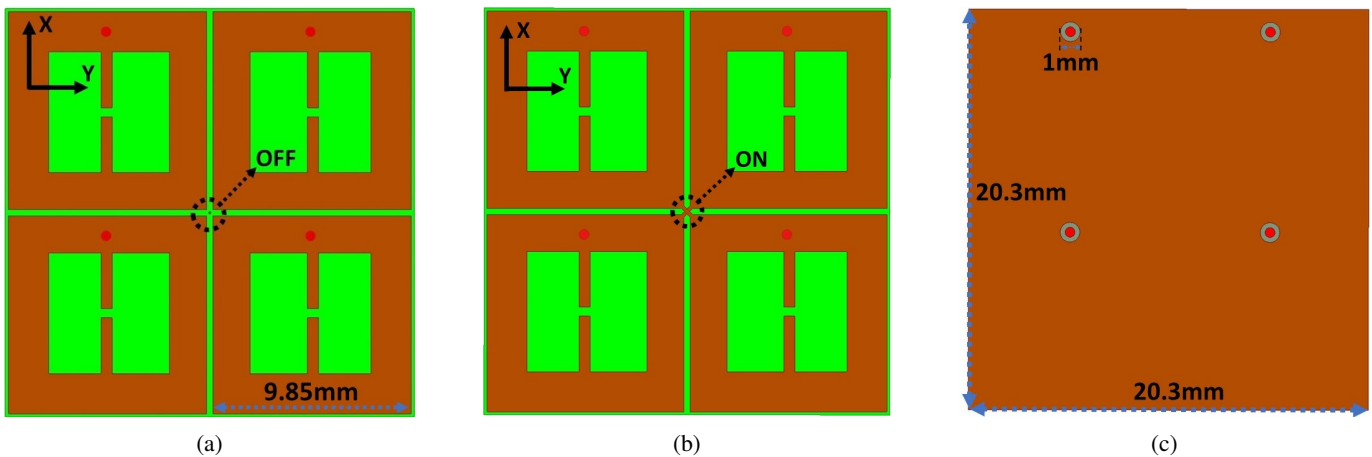

Fig. 8: Schematic view of the top surface of the proposed hybrid RIS supercell having the two states: (a) OFF; and (b) ON. (c) The bottom view of the proposed supercell. The red dots indicate copper, while the green color shapes represent structures of Rogers RO4003c material with a thickness of $0.813$ mm.

It should be noted that the impedance of each cell in the proposed supercell was designed at $150\,\Omega$; this parameter will be optimized in future research. Regarding the OFF state, it is demonstrated in Fig. 9a that, for normal incidence waves with linear polarization along the x-axis, $\eta_{RF-AC}$ of over $80\%$ is achieved at the frequency $4.7$ GHz. For y-polarized waves, the results showed a nearly ideal reflection (i.e., $S_{11}$ = -0.9 dB) at the frequency $5.05$ GHz. On the other hand, when the diode switches are short-circuited (i.e,, the ON state), the upper metallic layer establishes a connection with the ground plane through the central metallic via, which alters the frequency response of the unit cell. Specifically, absorption rates of $69\%$ at $5.5$ GHz and $89\%$ at $5.75$ GHz are achieved for a normal incident wave with polarization along the x- and y-axes, respectively, as shown in Fig. 9b. The results indicate that, for y-polarized normally incident waves, nearly ideal reflection is attained at the frequency $5.05$ GHz, and an $180$-degree phase difference in reflections between the ON and OFF states can be applied, as showcased in Figs. 9c and 9d, thus, enabling 1-bit reflection tunability.

We note that this paper introduces an initial conceptual framework for hybrid reflecting and RF energy harvesting RISs through a full-wave simulation study, which serves as the foundation for further optimization and the exploration regarding both the frequency and polarization sensitivity of the proposed hybrid meta-atom, as will be elaborated in the sequel. The proposed metasurface can offer a dual functionality, serving not only as an MEH, but also as a programmable reflector with two states, to favorably engineer wave propagation via various beamforming techniques adhering to specific communication goals [147]. Inspired from the aforedescribed initial full-wave simulation results of the designed supercell, in the following, we present two new RF energy harvesting policies, namely, frequency and polarization splitting.

### C. Policies for Energy Harvesting

The interplay between energy harvesting and reflection functionalities with the proposed hybrid RIS design requires a protocol to determine when and how each operation will take place. Such protocols can be found in the literature and are mainly categorized as time switching as well as power and element splitting [41], [42] and [40]. Time switching is quite straightforward and denotes the policy where time is divided into intervals of reflection and absorption (i.e., energy harvesting). Power splitting refers to the ability of the elements to store a percentage of the wave impinging on them, while reflecting the rest. Lastly, element splitting considers a division of the RIS elements, some to perform energy harvesting and others tunable reflections, benefitting from the harvested energy to power the reflecting elements. A comprehensive overview and comparison of these policies was carried out in [41], showcasing the advantages and disadvantages of each policy under different scenarios. In the following, we will describe the frequency and polarization splitting harvesting policies.

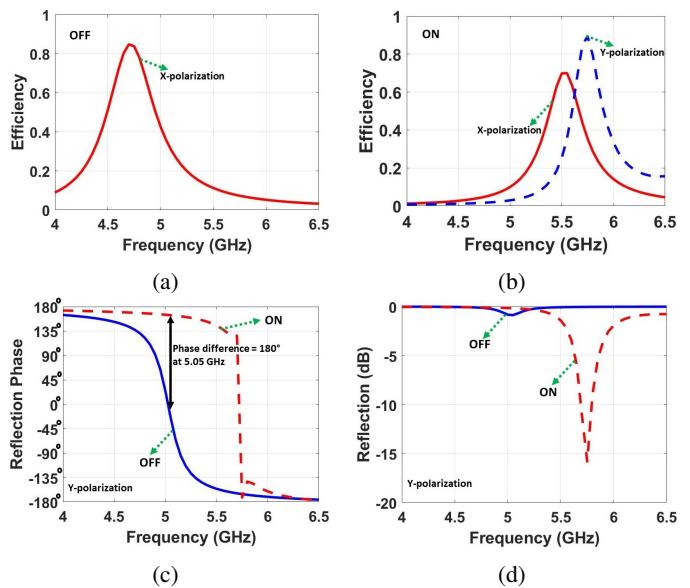

Fig. 9: Full-wave simulation results showcasing the performance of the proposed energy harvesting 1-bit RIS unit cell. The efficiency in the top two figures refers to the captured efficiency $\eta_{RF-AC}$ for: (a) OFF state with illumination by an x-polarized normal incident wave; and (b) ON state with illumination by both x- and y-polarized normal incident waves. Additionally, (c) reflection phase and (d) reflection level of the RIS in both ON and OFF states, when illuminated by a y-polarized normal incident wave, are illustrated.

*1) Frequency Splitting:* For this dual-functional policy, a harvesting band and a band of reflection influence are first defined. Then, each RIS unit cell is designed to achieve perfect absorption at the harvesting band (i.e., zero reflections), while a perfect reflection is desired at the band of influence. Ideally, for the different states of the element that are necessary for tunable reflections (for example, the ON/OFF states of PIN diodes), its frequency response in the harvesting band should remain stable to realize a reliable power supply. A first attempt to realize such an element is showcased in this paper, employing the harvesting unit cell proposed in [82]; this can be extended to other element designs. In future works, different frequency bands should be considered. Indicatively, the harvesting band could align with the bands of available ambient RF power from real-world measurements [148]–[150], while the band of influence could be specified for some 6G applications. Lastly, polarization insensitivity should be investigated as well to optimize performance.

*2) Polarization Splitting:* Polarization of an EM wave refers to the orientation of the electric field, which can be either linear, circular or elliptical [75]. For simplicity, we will address only transverse waves, i.e., linearly polarized waves where the electric field is perpendicular to the direction of propagation. As aforementioned, the proposed element in this paper can absorb and reflect x- and y-polarized waves, respectively (see Fig. 9). Inspired from this, we propose polarization splitting harvesting, where a metamaterial element can be designed with the objective of both harvesting and reflecting in the same frequency for orthogonally polarized waves. For example, vertically polarized waves could be used for energy harvesting and horizontal ones for tunable reflections (see Fig. 10); note that this assignment can change subject to the specific use case.

### D. Power Budget Evaluation

We will now investigate the feasibility of the proposed hybrid RIS design integrating MEH unit cells, focusing on the evaluation of the energy sustainability from ambient RF sources. We will first provide a brief description of the ambient RF power levels available in the Ultra High Frequency (UHF) spectrum (0.3-3GHz), and then, the power consumption modeling of available RIS designs will be reviewed. Finally, a feasibility study based on state-of-the-art advancements and experimental measurements on RIS power consumption and urban ambient RF power levels will be presented.

*1) Ambient RF Power Budget:* The UHF band is considered to be the most prominent for energy harvesting due to the high traffic over the whole band and low path loss compared to mmWave and beyond [151]. According to a study conducted within Greater London [148], the most available RF power was recorded in the GSM1800 band, spanning from $[1805, 1880]$ MHz, with an average power density of $84\,\text{nW/cm}^2$ (this value was obtained by adding all spectral peaks within this band). Similarly, in [149], the GSM1800 was proved to be the band with the most available power as well, but with significantly lower values, having a power density of $3.55\,\text{nW/cm}^2$, while the power density over the whole band within 680MHz and 3.5GHz was shown to be $5.62\,\text{nW/cm}^2$. In addition, in a more recent study in [150], ambient RF power was measured at the $[1850, 1860]$ MHz band in Nanjing, China to be $615\,\text{nW/cm}^2$; this value is significantly higher than the ones measured in [148], [149], considering also a shorter bandwidth. Another promising band for RF harvesting is the Digital Tele-Vision (DTV) band ranging in $[470, 700]$ MHz [152]. A distinct benefit of this band is its stability over time, since it does not depend on human activities, as the rest of the RF spectrum. From measurements conducted in [148], it was shown that the power values vary from $0.89$ to $469\,\text{nW/cm}^2$. Moreover, values for the power levels in the DTV band were given in [151, Fig. 3(a)], but with the unit of measurement being the electric field per meter ($E$) instead of the power per square meters. To this end, we convert the depicted values to power density using the formula $E^2/Z_0$ [153] with $Z_0 = 377\,\Omega$ being the free space impedance. We can then observe that, for the DTV bands in Tokyo, Japan and in Atlanta, USA, the power density per spectral peak was around $2.6\,\text{nW/cm}^2$ and $6\,\text{nW/cm}^2$, respectively. These values, if added for the whole DTV spectrum, can lead to significantly higher power density. Concluding, it is difficult to provide some general values regarding the power density over the entire UHF spectrum, since it is significantly location sensitive. In this subsection, an attempt was made to gather power spectral density values from real-world measurements, which are critical for the design of energy harvesting antennas in SREs.

*2) RIS Power Consumption:* The power consumption of an RIS depends on the hardware technology to implement its phase-tunable unit elements as well as its control mechanism, with the latter being responsible to sweep among the matasurface's reflection configurations and perform the interface with wireless network [45]. Both components are currently subject to extensive research by the relevant scientific communities, namely, the antenna, metamaterials, and wireless communications communities. We will next focus on RISs whose phase reconfigurability is enabled by electrical stimuli, actually serving as the actuation mechanism. We will specifically review the following metamaterial technologies: *i)* PIN diodes [144], [146], [154], [155]; *ii)* varactor diodes [144], [145], [156] diodes; and *ii)* RF-switches enabled by CMOS technology [143], [157], since most available prototypes and experimental results employ one of the three. The control unit that performs the computation tasks necessary for designing the response of each RIS element, as well as the driving circuits, play also a significant role in RIS's power consumption. Their cumulative energy requirements will be, henceforth, termed as the static power consumption of the RIS [144]. All in all, the RIS power consumption can be divided in the following three main

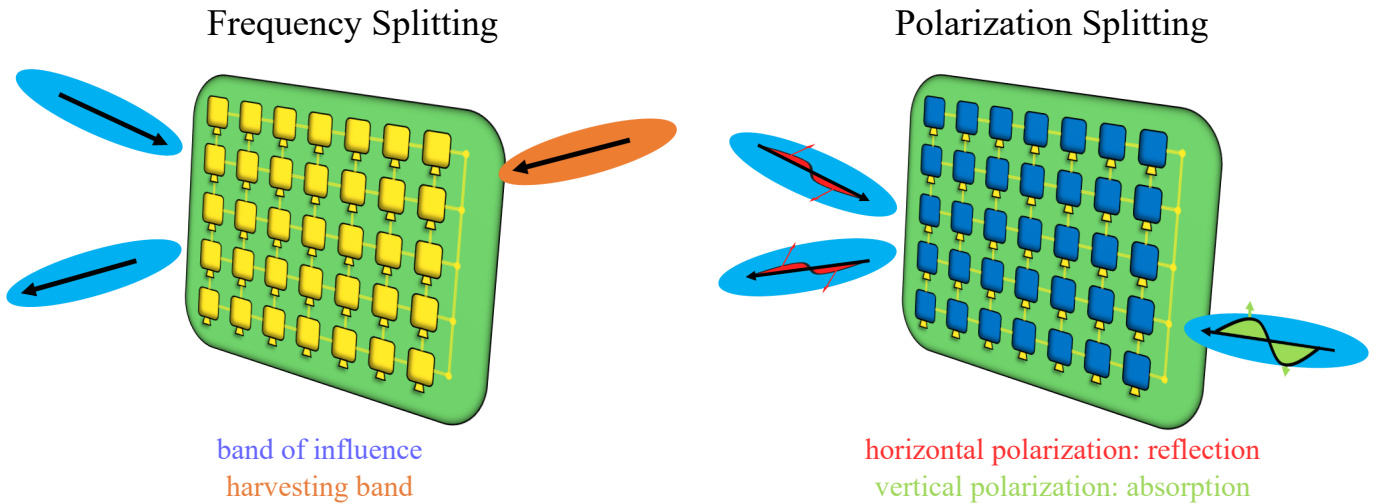

Fig. 10: Energy harvesting RIS with frequency splitting (left figure): the orange lobe indicates an impinging wave at the harvesting band, while the blue lobes depict an impinging wave and a reflected wave in the RIS band of influence [13]. Hybrid energy harvesting and reflecting RIS with polarization splitting (right figure): the red and light green arrows depict the horizontal and vertical polarizations of the impinging waves, which are exploited for reflection and harvesting, respectively.

categories:
- Power consumption of the RIS controller, $P_c$, which can be either an FPGA-based processor [144], [145], [155] or a microcontroller [143], [154], [156], [157], depending on the computational requirements. The latter can reduce power consumption from the order of W to mW [145].
- Power consumption of the driving circuits, $P_{dr,tot}$. This depends on the number of independently controlled phase-tunable elements, the switching device used in each of them, the polarization mode (single or dual), the phase shift resolution bits, $B$, and the number of control signals from each driving circuit [144].
- Power consumption of the RIS unit element, $P_u$, which depends on the enabling technology of each metamaterial. This power consumption value can also change with respect to each element's phase configuration. Indicatively, varactor diodes and RF-switches are configuration independent, while PIN diodes depend on the working state, i.e., the number of the ON and OFF diodes deployed. The value of $P_u$ also depends on the bit resolution $B$ (i.e., the number of phase/reflection states) and the polarization mode. Typically, to have dual polarizability, a pair of these basic devices needs to be integrated within each unit cell, i.e., for a 1-bit PIN diode unit cell with dual polarizability, two PIN diodes are necessary [144].

Based on the preceding definitions and by denoting with $N$ the number of RIS unit elements, the total power consumption of an RIS is given by the expression:

$$P_{\text{RIS}} \triangleq P_c + P_{dr,tot} + NP_u, \quad (4)$$

where each of these values depends on the specific hardware implementation. Indicatively, for unit cells based on RF-switches with single polarizability, $B$-bit phase shift resolution, and independent unit tunability, the power consumption of the driving circuitry is given as [144]: $P_{dr,tot} = BNP_{dr}$, where $P_{dr}$ is the driving power consumed for one unit cell. It is noted that $P_c$ and $P_u$ do not depend on the number of RIS elements and their phase resolution. On the other hand, for RISs based on PIN diodes, the power metric $P_u$ also scales with the phase resolution in bits, and generally, the unit cell power consumption is significantly higher than that of RF switches and varactor diodes. Furthermore, regarding varactor diodes, although they have negligible power consumption per unit element, their driving circuits consume a large amount of energy compared to RF-switches and PIN diodes due to the required digital-to-analog-converters, operational amplifiers, and level regulators.

Putting all above together, it can be concluded that, out of three aforediscussed RIS element technologies, the one based on RF switches constitute the most prominent technology in terms of power consumption, since it bears a good trade-off between the advantages of PIN and varactor diodes. To quantitatively illustrate this fact, some indicative values from the various available RIS prototypes are presented in[1] Table II. It is apparent that, since the first RIS prototypes developed in 2020 [155], power consumption has significantly decreased from 153 W to mW levels, and is expected to be reduced even further, given the extensive RIS research (see [46] and references therein). For example, the authors in [154] state that, by employing a microcontroller and RF switches, the RIS power consumption can be reduced to the orders of μW. It is finally noted that other technologies, such as MEMS, could potentially improve energy efficiency due to their non-volatile switching capability, since power will be needed only for the instance of reconfiguration and not for phase/reflection state maintenance.

*3) Energy Harvesting Feasibility Study:* We will now analyze the feasibility of powering conventional reflective RISs

---

[1] In Table II, the symbol "∼" indicates expected, and not actually measured, values.

TABLE II: Power consumption of available RIS hardware designs based on experimental measurements.

| Ref. | Unit Cell/ $P_u$ | $P_{dr,tot}$/ $P_{dr,tot}$ per unit cell | Number of unit cells $N$ | Freq. (GHz) | Phase shift resolution | Controller/ $P_c$ | $P_{\text{RIS}}$/ $P_c + P_{dr,tot}$ |
|---|---|---|---|---|---|---|---|
| [157] | RF switches HMC7992/ 0.47 mW | 3.96 mW/ 0.2475 mW | 16 | 5 | 2-bit | microcontroller STM32L053C6/ 2.31 mW | 13.86 mW/ 6.27 mW |
| [143] | RF switches SKY13418-485LF/ N/A | N/A | 100 | 5.3 | 3-bit | microcontroller STM32L071V8T6/ 62 mW | $\sim 62$ mW/ N/A |
| [144, (6#RIS)] | RF switches/ 0.495 mW | 240 mW/ 3.75 mW | 64 | N/A | 1-bit | FPGA XC7K70T/ 4.8 W | 5.072 W/ 5.04 W |
| [145] | Varactor diodes SMV2019-079LF/ 1.6 µW | 931.8 mW/ 0.847 mW | 1100 | 5.8 | 1-bit | FPGA Xilinx Zynq 7100/ 1.5 W | 2.434 W/ 2.432 W |
| [156] | Varactor diodes/ 5 fW | N/A | 984 | 5.25 | continuous | microcontroller | $\sim 1$ W/ $\sim 1$ W |
| [158] | Varactor diodes/ N/A | N/A | 256 | 4.25 | continuous | FPGA | 0.7 W/ N/A |
| [144, (5#RIS)] | Varactor diodes/ $\sim 0$ mW | 1.72 W/ 13.4 mW | 128 | 3.2 | continuous | FPGA/ N/A | $\sim 6-7$ W/ $\sim 6-7$ W |
| [159] | PIN-diodes (dual-polarized)/$\sim$ 1.9 mW (ON state) | N/A | 1600 | 28 | 1-bit | FPGA/ N/A | 16 W (all diodes at ON state)/ 13 W |
| [144, (2#RIS)] | PIN-diodes (dual-polarized)/ 25.12 mW (ON state) | N/A | 3600 | 35 | 1-bit | FPGA/ N/A | 103.2 W (all diodes at ON state)/ 15.75 W |
| [144, (3#RIS)] | PIN diodes/ 11.99 mW (ON state) | N/A | 512 | 2.6 | 1-bit | FPGA/ N/A | 12.66 W (all diodes at ON state)/ 6.52 W |
| [155] | PIN diodes/ N/A | N/A | 256 | 2.3 | 2-bit | FPGA/ N/A | 153 W (all diodes at ON state)/ N/A |
| [154] | PIN diodes/ 3 mW (ON state) | 80 mW/ 0.5 mW | 160 | 5.8 | 1-bit | microcontroller Arduino MEGA2560/ 100 mW | 660 mW (all diodes at ON state)/ 180 mW |

via energy harvesting from ambient RF sources. To conduct this feasibility study, the amount of potential harvested energy needs to be first characterized. To do so, we will focus on the following metrics: the spectral ambient RF power density $\mathcal{S}(f)$ in W/Hz/m², the RF-to-DC harvesting efficiency $\eta_{RF-DC}$, the antenna's aperture $A$, and the total bandwidth $BW_H$ used for energy harvesting. Assuming polarization insensitivity and that $\eta_{RF-DC}$ is constant over $BW_H$, the amount of EM incident power transformed directly to DC power, $P_{DC}$, can be expressed as follows: [150]

$$P_{DC} = \eta_{RF-DC} P_{RF}, \quad (5)$$

where the incident EM power on the RIS panel, $P_{RF}$, is given by the expression:

$$P_{RF} = A \int_{BW_H} \mathcal{S}(f) df. \quad (6)$$

In the feasibility analysis, the measurements from [150] will be considered, since to the authors' best knowledge, it is the most recent study providing ambient RF power levels with respect to the antenna aperture. Therein, an approximated expression for $\mathcal{S}(f)$ is provided for the spectrum range [1850, 1860] MHz [150, eq. (4)]. Note that, according to [150, Fig. (10)], the ambient power levels are nearly constant in the [1820, 1900] MHz band. Consequently, we assume the RF energy harvesting bandwidth to be $BW_H = [1820, 1900]$ MHz and, to get a rough measurement for the power density in this spectrum, [150, eq. (5)] is linearly extrapolated in this range, yielding:

$$\frac{P_{RF}}{A} = \int_{BW_H} \mathcal{S}(f) \, df \approx 4.92 \mu\text{W/cm}^2. \quad (7)$$

For similar power density levels, specifically 2.6µW/cm², a nearly 100% efficiency is achieved with respect to the physical aperture of the MEH in [92], as previously described in Section III. Subsequently, based on these results, an optimistic estimate of the harvested power density per physical antenna surface, $P_{DC}/A$, can be evaluated. By combining (5)-(7) and invoking $\eta_{RF-DC} = 1$ (ideal harvesting), it yields that[2] $P_{DC}/A = 4.92$ µW/cm². Finally, the aperture of the harvesting surface needed to power an RIS can be calculated as $A = P_{\text{RIS}}/4.92\,\mu\text{W cm}^2$. For purposes of completeness, the general formula needed to calculate $A$ is:

$$A = \frac{P_{\text{RIS}}}{\int_{BW_H} \eta_{RF-DC}(f)\mathcal{S}(f)\,df}. \quad (8)$$

---

[2] It is noted that these power values depend on the specific RIS element design, and to achieve simultaneous reflection and energy harvesting, a compromise between harvesting efficiency and desired reflection configuration might be needed, even if these two functionalities happen in separate bands (i.e., frequency splitting); this is an interesting topic for further investigation.

The latter expression can be straightforwardly used to compute the aperture of the harvesting surface/antenna needed to power an RIS, providing valuable insights on the RIS designs that could be potentially powered by ambient RF sources.

We commence by studying the most power consuming component of an RIS, which is its control board [160]. In the recent prototypes in [144]–[146], an FPGA processor was used for the RIS controller, which has a power consumption in the order of W. Typically, the power consumption of the controller is not affected by the number of RIS elements, but since the complexity of the algorithms that should be implemented usually scales with the number of elements, the power consumption might scale as well. As a result, low complexity algorithms should be developed as in [145], where even with 1100 elements the power consumption of the FPGA was 1.5 W. Nevertheless, even if that optimistic assumption of 1.5 W power consumption is made, that would require, using the formula in (7), a $30.49\,\text{m}^2$ ideal antenna surface to power the controlling board alone. It can thus be easily seen that such processing hardware is too power-hungry for ambient RF energy harvesting, implying that one would have to resort to microprocessors for implementing the RIS controller, which seem to be a more feasible solution. Specifically, in [143], the power consumption of the microprocessor unit was measured to be 62 mW, which could be reduced to the order of μW if low power consuming modes are considered. Moreover, as shown in Table II for a small RIS consisting of 16 elements, the microcontroller consumes only 2.31 mW.

Another important aspect regarding the power consumption of an RIS is the selection of unit cell hardware. At first glance, by inspecting Table II, the varactor diode elements seem to be the best choice. However, all factors considered, their cumulative power consumption is at the order of W due to the required power-hungry driving circuitry, as elaborated in Section V-D2. As a result, for our feasibility analysis, we will focus on the RF switch-based RISs due to their two following major advantages. Firstly, as previously illustrated, they are the most promising energy-wise, and secondly, they are capable of signal absorption on top of phase-tuned reflections, implying strong potential for energy harvesting [143]. Following [157], the measured power consumption $P_{RIS}$ for an RIS consisting of RF switches and a microcontroller is given in Table II. We now list the two hybrid RIS designs fow which we will conduct our energy harvesting feasibility analysis:

- Design 1: Hybrid RIS with frequency splitting and ideal harvesting at the frequency band of $BW_H = [1820, 1900]$ MHz from ambient RF sources; the reflection band of influence is considered to be centered at the frequency $f_c = 5$ GHz [157].
- Design 2: Hybrid RIS with both frequency and element splitting harvesting, i.e., all RIS elements will perform harvesting simultaneously (frequency splitting as in Design 1), but only a portion of them will have reflective capabilities (element splitting).

For both designs, to relate the hybrid RIS's aperture with the number of reflective elements, each unit cell is considered to have a size of $\lambda/2 \times \lambda/2$ with $\lambda$ being the operating wavelength of influence for reflections of the impinging signals.

It can be trivially concluded for Design 1 that the harvesting power per unit cell, $P_{DC,u}$, needs to be at least larger than the cell's average power consumption (i.e., $P_{DC,u} \geq P_{dr,tot}/N + P_u$), so that, by scaling to more elements, the RIS controller will be powered from the excess energy. For the considered power requirements in the first row of Table II, it needs to hold that $P_{DC,u} \geq 0.7175$ mW. To ensure that this condition is met, the area of each unit cell, $A_u$, needs to satisfy the inequality $A_u \geq 145.83\,\text{cm}^2$. Given that $A_u$ and $\lambda$ are related as $A_u = \lambda^2/4$, yields $\lambda \geq 24.15$ cm implying that $f_c \leq 1.24$ GHz. It can thus be concluded that, with the current advances in RIS technology for Design 1, it is not possible to realize a sustainable RIS in terms of energy with only ambient RF energy harvesting and frequency splitting. It must be noted, however, that the available prototypes in Table II are not optimized in terms of power consumption. It is expected that, with the ongoing and future advances on the RIS technology [135], the power consumption values of their components will drop to μW levels.

For an RIS designed to provide programmable reflections at 5 GHz to be energy autonomous, the power consumption per element (i.e., $P_{dr,tot}/N + P_u$) would have to be at maximum $44.2\,\mu\text{W}$, which is believed to be achievable [143], [154], [157]. For sub-THz and THz frequencies, other materials should be considered, such as liquid crystals, MEMs, and microfluids [106], [141], [161], which are even less power consuming. For instance, at $f_c = 100$ GHz, the maximum power consumption per element should be $0.11\,\mu\text{W}$, which might indicate that frequency splitting is not a viable solution at higher frequencies. On top of that, dual-band operation at both UHF and THz frequencies would probably be problematic, if not impossible, due to the large frequency difference. The latter issue could potentially be solved with element splitting. Specifically, one MEH operating at UHF for energy harvesting could be employed to power a physically smaller THz RIS intended for tunable reflections.

For Design 2, there is no restriction regarding the power consumption per RIS element since element splitting is employed. To demonstrate the feasibility of such an energy sustainable RIS, we will now characterize the aperture needed to power the RIS reported in [157], and give the ratio of the harvesting surface ($A$) to the reflecting surface, $A_R$. To this end, $A$ can be calculated via (7) and (8) by invoking the value $P_{\text{RIS}} = 13.86$ mW, which yields $A = 0.282\,\text{m}^2$. Note that $A_R$ is nearly 20 times smaller than $A$ for operation at 5 GHz (i.e., $A_R = 0.0144\,\text{m}^2$). These values are to be interpreted in a conceptual, and not a strict, manner, i.e., the fact that a 20 times larger MEH (for the ideal RF energy harvesting case) is needed to power a reflective RIS consisting of RF switches and a microcontroller implies that there is still a lot of room for innovation towards efficient hybrid RIS designs with energy autonomicity.

*4) Concluding Remarks:* Based on the previous energy harvesting feasibility study capitalizing on the available RIS prototypes summarized in Table II, we now sketch the relevant take-out messages along with some remarks on the potential of the available harvesting policies, namely, frequency splitting, power splitting, element splitting, and time switching).

- Ambient RF energy harvesting should be employed in the UHF frequency band, where there is rich spectrum occupancy with wide coverage. We have specifically identified the $[1805, 1880]$ MHz (GSM1800) and $[470, 700]$ MHz (DTV) bands [148] as the most promising. Especially for the DTV spectrum, which is only slightly affected by human activity, and thus, serves as a more stable and reliable source of ambient power, more experimental measurements need to be conducted to empirically characterize its power spectral density $\mathcal{S}(f)$.
- With the considered reflective RIS consisting of RF switches and a microcontroller and ambient RF power levels of around $5$ μW/cm$^2$, it is very difficult to realize an energy autonomous hybrid RIS integrating that reflective RIS with an MEH (for both Designs). Indicatively, for Design 1 the RIS power consumption level needs to drop down to $40 - 50$μW per unit cell for sub-6 GHz and lower than $0.1$ μW for sub-THz frequencies, for the hybrid RIS Design 1 to be feasible in an efficient way.
- The proposed hybrid RIS realizing frequency splitting (Design 1) for energy harvesting can be a viable solution for sub-6 GHz frequencies, if the power consumption of its reflective part (RIS panel and microcontroller) drops more than one order of magnitude compared to [157].
- A hybrid RIS with power splitting is not a feasible solution for energy autonomicity, since, even at the extreme case of only impinging signal absorption and consequent ideal harvesting, the harvested energy does not suffice.
- Polarization splitting is suboptimal for ambient RF energy harvesting due to the fact that impinging waves are mostly high-order bounced reflections with unknown polarization a priori [162]. A polarization-insensitive harvester would make more sense, otherwise, the power density levels that can be harvested will decrease even further. However, if a dedicated power beacon wirelessly transfers energy to the hybrid RIS with a known polarization, then polarization splitting would be a strong candidate for realizing an energy sustainable RIS.
- The proposed hybrid RIS Design 2 with element splitting provides another feasible energy sustainable solution. An extreme case could be to completely distinguish in the hybrid RIS architecture the part for energy harvesting (realized via a dedicated MEH) from that for programmable reflection (this RIS part will be powered by the MEH).
- Although time switching was not studied for a hybrid RIS, it can be intuitively concluded that it will face the same challenges as frequency and power splitting. In addition, even when a reflective RIS is idle (tunable reflections deactivated), the controller consumes some static power. This implies that, in the harvesting-only intervals of time switching, the RIS will consume some non-negligible power. Even if it is feasible for an energy harvesting hybrid RIS to be powered with time switching, most of the time with the available RIS prototypes would be spent on harvesting, and little-to-no time will be left for programmable reflections.
- It is finally noted that, to conduct a more rigorous analysis, polarization mismatches and non-ideal harvesting efficiencies need to be considered. Even though an ideal harvesting efficiency $\eta_{RF-DC}$ was reported in [92], this is generally not the case, especially for power density levels lower than 20 μW. This happens due to the rectifying circuitry typically underperforming at this regime [90].

## VI. Open Challenges And Research Directions

### A. Next Generation MEH Designs

The design of multiband and wideband MEHs remains an open challenging research direction. Further research efforts are needed to model an appropriate impedance-matching network between the MEH and its rectifying circuitry operating at multiple frequency bands. Recent experimental research on MEH typically employ a high-gain antenna with substantial output power to activate the rectifying diodes on the harvesting metasurface, while the intensity levels of ambient RF signals in real-world environments are much less, as discussed in detail in [47], [48], [92], [163], [164]. Another challenging research issue with future wireless networks is the lack of knowledge regarding the amount of power required for supplying low-powered wireless devices and IoT, as well as the amount of ambient RF energy available in the envisioned dense network deployment. To address this problem, wireless communication bodies and industries should introduce several technical standards in the field of energy harvesting and MEH measurements.

A compact MEH design with a high capturing efficiency may have multiple wireless applications, e.g.: *i)* energy harvesting of ambient RF signals; *ii)* WPT; and *iii)* receiving wireless information and energy simultaneously [47], [48], [163], [164]. Furthermore, advances in MEH in terms of high-efficiency levels even at oblique incident angles, miniaturization, flexibility, and simplicity in practical designs are needed. To this end, utilizing textile substrates with mechanical capabilities to bend and twist offer promising opportunities for improved wireless powering of electronic devices located in hard-to-reach areas of wireless environments.

### B. Energy Sustainable Reflective RIS

The proposed designs for hybrid RISs aiming at programmable reflections with energy autonomicity constitute a first attempt towards energy sustainable RIS-enabled SREs. The optimization of their unit elements as well as their harvesting, phase tunability, and computation components will open up various research directions, such as for simultaneous ideal reflections and harvesting, multi-bit reflection tunability with harvesting, and dual-band harvesting and reflection in the case where the two functionalities take place orthogonally in frequency. Moreover, new RIS prototypes targeting energy consumption minimization need to be investigated in order for this technology to be ready for large-scale deployment. To this end, RIS designs having the ability to be completely passive when turned off can be another future direction. Specifically, with sustainability and energy-efficiency being crucial for their development, it will be imperative to efficiently conserve energy when it is not needed and that the RIS can enter partial or fully power-saving sleep-mode when, for example, data traffic in its proximity is low.

### C. Energy Sustainable Simultaneous Reflective and Sensing RIS

A hybrid active and passive RIS relying on the metaatom design proposed in [165], which is capable of realizing simultaneous programmable reflection and absorption of the impinging signals, has been recently presented in [166]. As illustrated in Fig. 11, each $n$-th hybrid meta-atom has the ability to sense a portion of the impinging wave $(1 - \rho_n)$, possibly weighted through a tunable analog coefficient $\psi_n$, and reflect the rest of it ($\rho_n$) with a desirable tunable phase shift $\phi_n$. These meta-atoms are either individually or groupconnected to one or more reception RF chains, thus, bringing the captured incoming signal to a baseband unit, embedded to the RIS structure, for further processing [167]. The sensing capability of this hybrid RIS design is expected to facilitate various network management functionalities, including channel parameter estimation [168] and localization [169], contributing towards computationally autonomous and self-configuring metasurfaces with minimal control signaling [170], [171]. An interesting research direction deals with the energy sustainability, either partially or fully, of this hybrid RIS paradigm, which will efficiently address the additional power supply required, as compared to solely reflective RISs lacking reception RF chains and a baseband processing unit.

To achieve reconfigurability simultaneously in the phase profile (i.e., programmable reflection) and the power coupling (i.e., spatially tunable sensing/reception), a more accurate power consumption modeling is required for the tuning mechanisms and the relevant driving circuits. Additionally, the power requirements for the reception RF chains alone exceed the values that can be compensated by RF energy harvesting from ambient sources. Indicatively, one reception RF chain typically comprises a low noise amplifier, a mixer, a local oscillator, a low-pass filter, a baseband filter, and an Analogto-Digital Converter (ADC). Some indicative values for the power consumption of each of the latter components are given in [172], according to which, their cumulative power consumption can be around 260 mW. Towards reducing the power consumption of RIS hardware architectures capable of signal absorption and reception, it was proposed in [173] and [174] to employ 1-bit ADCs, which can drastically bring the power consumption of the reception RF chain to 62 mW [173]. However, this is still considered prohibitive for ambient RF energy harvesting. A more viable solution would be to employ amplitude detectors as proposed in [175] and [176], which can be easily integrated to the concept of energy harvesting RISs. With these detectors being embedded to the hybrid reflective and sensing RIS, the impinging signal will be again divided between reflected and absorbed signal, but for the absorbed signal, there will be two options: *i)* energy harvesting and *ii)* amplitude detection, as shown in [175, Fig. 1]. This architecture is promising since it eliminates the need for reception RF chains. However, the exact power consumption and working principle of such a computationally autonomous, self-configuring, and energy sustainable RIS design are still under investigation, and further research is needed to conclude if they can be completely or partially powered by energy harvesting via power splitting. Concluding, hybrid active and passive RISs with a single [6], or very few, reception RF chains consisting of minimal resolution ADCs (even 1-bit ones) and amplitude detectors for sensing deserve further studies in the context of energy sustainability for realistically assessing the potential of being powered through RF energy harvesting, either from ambient and/or dedicated power sources.

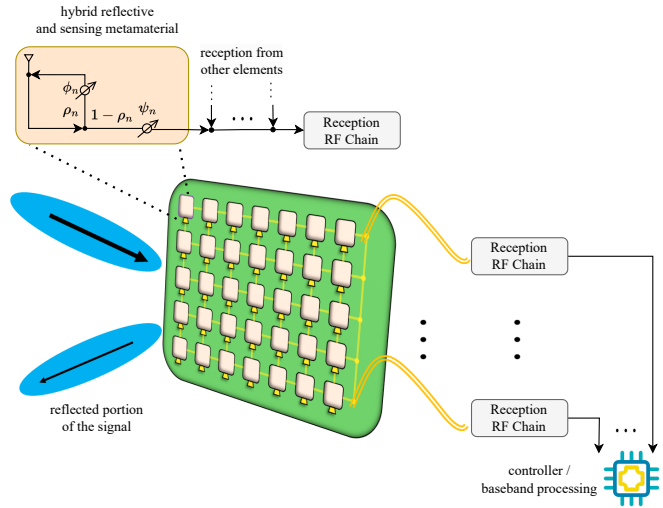

Fig. 11: The working principle of the hybrid active/passive RIS relying on [165]'s meta-atom design that realizes simultaneous tunable reflection and absorption of the impinging wave [166]. The absorbed portion of the impinging wave is guided to a single or multiple reception RF chains for sensing purposes via a dedicated baseband processing unit (can be integrated with the conventional RIS controller [160]).

### D. Towards the Sustainability of Multi-Functional Metasurfaces

Capitalizing on the recent advances in metametamaterial technologies for RISs in various frequency bands that offer efficient reconfigurability in ultra low power consumption manners and the inherent scalability feature of RISs, there is lately increasing interest in new RIS hardware architectures [15] enabling improved and/or new operation modes. The most prominent of them are:

- *Active RIS*: These metasurfaces include active components for amplifying the reflection of impinging waves, either via individually connected negative resistance components, such as tunnel diodes, for low-power amplification [177], or in an amplify-and-forward manner [178], where two RISs are interconnected with a single variable-gain amplifier. Their main advantage lies in their low cost reflection amplification, avoiding the need to convert the impinging signal to the digital domain, since they do not perform signal reception.

- *Stacked Intelligent Metasurface (SIM)*: This concept considers multiple RISs placed very closely and parallel to each other, with each RIS being spatially fed from the radiation of the previous one, as depicted in [15, Fig. 4]. Consequently, SIM comprises a 3D slab of RISs, which are reconfigured via one dedicated controller that is connected to all of them. The advantage of an SIM lies in its resemblance to an artificial neural network [179]. To this end, SIM structures is envisioned to be able to perform computationally intensive tasks, such as image and pattern recognition, entirely in the analog domain [180].

- *Simultaneous Transmitting and Reflecting (STAR) RIS*: These RIS structures are capable of simultaneous programmable refraction (i.e., retransmission) and reflection of the impinging signal at the reflective side of the metasurface, which can theoretically provide $360°$ signal coverage [181]. This unique characteristic enables network flexibility since there will no space restrictions for the deployment of BSs and end users. Consequently, STAR RISs can offer connectivity extension from two physically disconnected spaces by employing them on windows, walls, etc. [13]. However, an efficient hardware design for these architectures does not yet exist. Conceptually, each element has a transparent substrate and composes of a parallel resonant LC tank and small metallic loops to provide the required electric and magnetic impedance in both sides of the metasurface.

- *Dynamic Metasurace Antenna (DMA)*: This emerging paradigm, refers to metasurface-based transceiver architectures [182]. Their RF front end comprises a planar metasurface where metamaterial elements are placed in row-(or column-)groups to compose a microstrip antenna similar to linear antenna array. A DMA consists of multiple microstrip antennas that are individually connected to an RF chain, either for transmission or reception. In this manner, both analog and digital beamforming can be realized for improved performance. Their distinct and valuable feature is that they can facilitate the envisioned extremely massive Multiple-Input-Multiple-Output (MIMO) communications by employing multiple meta-element radiators, while keeping power consumption low compared to conventional massive MIMO [19], [182]. A recent DMA proposal for reception includes 1-bit ADCs per microstrip antenna to substantially decrease overall power consumption, while profiting from the large metasurface panel to compensate the quantization-induced loss in rate performance [173].

Although all the aforementioned RIS designs are typically considered as low power consuming (some including almost passive and others active components), rigorous power consumption models as well as hardware prototypes are still missing from the literature. In addition, future research could be conducted to investigate their integration with an MEH on either a simulation or an experimental basis, and the viability of RF energy harvesting to power these RIS hardware architectures, making fully or partially energy sustainable.

## VII. CONCLUSIONS

In this survey, state-of-the-art advances regarding the MEH technology were assembled and the fundamentals of MEHs were illustrated using both transformer and impedance network equivalents. Furthermore, 2D isotropic MEHs were investigated for achieving increased angular stability, which is of paramount importance for ambient RF energy harvesting. Thereon, we delved into RISs, describing the available unit cell designs and gathering current prototypes and experimentally measured power consumption levels. A novel hybrid RIS unit cell was introduced with the ability of RF energy harvesting on top of 1-bit tunable reflection, which provides a conceptual framework for future research and optimization. Motivated by our findings, we researched the available RF ambient power levels and conducted a realistic feasibility analysis regarding an energy autonomous integrated MEH-RIS structure. Capitalizing on that analysis, we deduced significant remarks regarding the systematic setup that can facilitate next generation energy sustainable hybrid RISs, such as the size limits of such a metasurface, its metamaterial technology, the frequency bands of operation, and the energy harvesting policies employed. Lastly, we identified future research directions concerning MEH, energy sustainable hybrid RISs, metasurface-based transceivers, and the interplay between the latter and RF energy harvesting, towards energy sustainable 6G, and beyond, networks.